\documentclass[journal]{IEEEtran}
\usepackage{amsfonts}
\IEEEoverridecommandlockouts

\ifCLASSINFOpdf
\else
\fi
\usepackage{proof}
\usepackage{multirow}
\usepackage{cite}%
\usepackage{stfloats}
\usepackage{color}
\usepackage{epsfig}
\usepackage{graphicx}%
\usepackage{psfig}
\usepackage{epsf}
\usepackage{slashbox}
\usepackage{float}
\usepackage{amssymb }
\usepackage[cmex10]{amsmath}
\newtheorem{remark}{Remark}

\begin{document}
\title{System Outage Probability and Diversity Analysis of SWIPT Enabled Two-Way DF Relaying under Hardware Impairments}
\author{\IEEEauthorblockN{Guangyue Lu, Zhipeng Liu, Yinghui Ye,~\IEEEmembership{Member,~IEEE}, and Xiaoli Chu,~\IEEEmembership{Senior Member,~IEEE}} %\vspace{-10pt}\Delta
\thanks{ Guangyue Lu, Zhipeng Liu, and  Yinghui Ye  are with the Shaanxi Key Laboratory of Information Communication Network and Security,  Xi'an University of Posts \& Telecommunications, China. (e-mail: tonylugy@163.com, zhipeng\_liu\_steve@163.com, connectyyh@126.com)}
 \thanks{Xiaoli Chu  (x.chu@sheffield.ac.uk) is with the Department of Electronic and Electrical Engineering, The University of Sheffield,  U.K.}
%\thanks{This work was supported by the Science and Technology Innovation Team of Shaanxi Province for Broadband Wireless and Application (2017KCT-30-02), and the Postgraduate Innovation Fund of Xi'an University of Posts \& Telecommunications (CXJJLZ2019026).}
}
\maketitle
\begin{abstract}
This paper investigates the system outage performance of a simultaneous wireless information and power transfer (SWIPT) based two-way decode-and-forward (DF) relay network, where potential hardware impairments (HIs) in all transceivers  are considered.\;After harvesting energy and decoding messages simultaneously via a power splitting scheme, the energy-limited relay node forwards the decoded information to both terminals.\;Each terminal combines the signals from the direct and relaying links via selection combining.\;We derive the system outage probability under independent but non-identically distributed Nakagami-$m$ fading channels.\;It reveals an overall system ceiling (OSC) effect, i.e., the system falls in outage if the target rate exceeds an OSC threshold that is determined by the levels of HIs. Furthermore, we derive the diversity gain of the considered network. The result reveals that when the transmission rate is below the OSC threshold, the achieved diversity gain equals the sum of the shape parameter of the direct link and the smaller shape parameter of the terminal-to-relay links; otherwise, the diversity gain is zero.
This is different from the amplify-and-forward (AF) strategy, under which the relaying links have no contribution to the diversity gain. 
Simulation results validate the analytical results and reveal that compared with the AF strategy, the SWIPT based two-way relaying links under the DF strategy are more robust to HIs and achieve a lower system outage probability.
\end{abstract}
\vspace{10pt}
\begin{IEEEkeywords}
Decode-and-forward relay, diversity gain, hardware impairments, simultaneous wireless information and power transfer, system outage probability.
\end{IEEEkeywords}

\section{Introduction}
Owing to its high  spectrum efficiency, two-way relay networks (TWRNs), which achieve bidirectional message exchange between two terminals via a relay node, are gaining significant momentum in the area of Internet of Things \cite{5273815}. However, the development of TWRNs is facing challenges, particularly due to the limited battery capacity of the relay node \cite{6552841}.To address this issue, simultaneous wireless information and power transfer (SWIPT) technique has been integrated into TWRNs, which can be applied in wireless sensor networks to enhance the communication quality between sensors. The key idea of SWIPT based TWRNs is to allow the energy-limited relay node to harvest energy from incident radio frequency (RF) signals through either a time switching (TS) or power splitting (PS) scheme, and  use the harvested energy to assist the transmission between two terminals \cite{6957150}. The design of PS or TS schemes and the performance analysis for SWIPT based TWRNs have been studied  \cite{7782410,7807356,8633928,7831382,8653432,8364583,7847314,8556567,8613860,8576644,8361446}.

%In \cite{7807356}, the authors considered a TWRN under a multiple access broadcast (MABC) protocol,
%%\footnote{In MABC protocol, there need two time slots to accomplish message interchange between two terminals, but the direct link between the two terminals can not be utilized due to the half-duplex constraint \cite{7831382}.},
%and jointly optimized the TS/PS and time phase (TP) ratios to minimize the system outage probability.

For a PS-SWIPT enabled two-way decode-and-forward (DF) relay network under a time division broadcast (TDBC) protocol, the authors in \cite{7782410} studied the terminal-to-terminal (T2T) outage performance. Shi \emph{et al.} \cite{8633928} considered a two-way DF relay network under a TDBC protocol, and proposed a dynamic PS scheme to enhance system outage performance.
%For a two-way decode-and-forward (DF) relay network under a time division broadcast (TDBC) protocol,
%%\footnote{In TDBC protocol, The direct link can be utilized, but the information interchange needs three time slots \cite{8576644}.}
%Shi \emph{et al.} \cite{8633928} proposed a dynamic PS scheme to enhance system outage performance.
Considering a SWIPT based two-way DF relay network, the authors in \cite{8653432} investigated the tradeoff between the achievable data rate and the residual harvested energy at the relay node.
In \cite{7847314}, the authors considered a SWIPT enable cognitive TWRN under a multiple access broadcast (MABC) protocol, and jointly optimized the PS ratio and interference temperature apportioning parameters to maximize the achievable throughput.
For a two-way DF relay system under a TDBC protocol, the authors in \cite{8613860} derived the system outage probability, and then studied the influence of various parameters on system outage performance.
Ye \emph{et al.} \cite{8361446} considered a SWIPT based two-way multiplicative amplify-and-forward (AF) relay network under a TDBC protocol, in which a dynamic asymmetric PS strategy was proposed to optimize the system outage performance.
%In \cite{8613860}, the authors studied the system outage performance of a two-way DF relay system under a TDBC protocol.
Nonetheless, all the above works assumed ideal hardware, but in practical system, all transceivers always suffer from a variety of hardware impairments (HIs), e.g., quantization error, inphase/quadrature (I/Q) imbalance and phase noise \cite{6214150,6630485,9437961,9239335}.
Despite sophisticated mitigation algorithms, the residual HIs may still have a deleterious impact on the achievable performance \cite{6630485}. %Thus it is necessary to study the impacts of HIs in the SWIPT systems.

Several recent works on SWIPT based TWRNs have taken HIs into account \cite{7247469,7835665,8851300,1234567}. For two-way cognitive relay networks, the authors in \cite{7247469,7835665} studied the impact of HIs on the T2T outage performance with a TS or PS scheme respectively.
%\textcolor{blue}{For a TS-SWIPT based two-way hybrid decode-amplify-forward (HDAF) relay network with a MABC, the authors in \cite{8851300} derived the T2T outage probability under HIs, and then proposed a improving HDAF relaying strategy to enhance the outage performance.}
For a TS-SWIPT based TWRN under a MABC protocol, the authors in \cite{8851300} considered HIs, and studied the T2T outage performance.
%In \cite{8851300}, the authors considered a TS-SWIPT based two-way hybrid-decode-amplify-forward (HDAF) relay system under a MABC protocol, and derived the T2T outage probability in the presence of HIs.
The results in \cite{7247469,7835665,8851300} show that HIs deteriorate the T2T outage performance, particularly in high-rate transmissions. Recall that the system outage probability, which jointly considers the outage evens of both terminals and the correlation between the two links, is also critical to designing practical systems.
%Please note that, the above works \cite{7247469,7835665,8851300} mainly focuses on T2T outage performance that considers only the outage event of one terminal. However, quantifying the system outage probability, which jointly considers the outage evens of both terminals and the correlation between the two links, is critical and is much more challenging than the T2T outage probability.
%The T2T outage probability considers only the outage event of one terminal, while the system outage probability jointly considers the outage evens of both terminals, as well as the correlation between the two links. Accordingly, quantifying the system outage probability is critical and is much more challenging than the T2T outage probability.
The authors in \cite{1234567} studied the system outage performance for a SWIPT based two-way AF relay network under HIs. Their theoretical analysis and simulations revealed that the relaying links contribute zero diversity gain because the AF strategy amplifies the signals and distortion noises simultaneously.
On the contrary, the DF strategy has the advantage of eliminating noise accumulated at the relay node, and thus the distortion noise amplification can be avoid. Based on the above observations, a natural question arises:  can this property alleviate the impact of HIs and bring performance gains in terms of the system outage probability and the diversity gain by comparing it with the AF strategy?

In this paper, we aim to answer the above questions. More specifically, we consider  a PS-SWIPT based two-way DF relay network under a TDBC protocol$^1$\footnotetext[1]{Compared with MABC, TDBC enjoys a lower operational complexity at the relay node and utilizes the direct link \cite{8613860,8576644}, hence we adopt the TDBC instead of the MABC.}, where HIs at all transceivers are considered, and investigate the system outage probability$^2$\footnotetext[2]{Note that \cite{8851300} focused on the T2T outage probability, which counts the outage event of one terminal only and is different from the system outage probability considered in our work.} and the achievable diversity gain.
\begin{table}[t]
\caption{Notations in this paper}
\centering
\begin{tabular}{|c|c|} %l(left)居左显示 r(right)居右显示 c居中显示
\hline
Notation&Definition\\
\hline
$\mathcal{CN}\left( {a,b} \right)$&Gaussian random variable with mean $a$\\
  & and variance $b$\\
\hline
${\rm{Naka}}(c, d)$&Nakagami random variable with fading \\
  &severity parameter $c$ and average \\
  & power $d$\\
\hline
$\Pr \{  \cdot  \}$&Probability of an event\\
\hline
$\left|  \cdot  \right|$&Absolute value of a number\\
\hline
$f_X(x)$&Probability density function (PDF)\\
\hline
$F_X(x)$&Cumulative distribution function (CDF)\\
\hline
$\Gamma(\cdot)$&Complete gamma function\\
\hline
$\gamma \left( {\cdot,\cdot} \right)$&Lower incomplete gamma function\\
\hline
$\Gamma \left( {\cdot,\cdot} \right)$&Upper incomplete gamma function\\
\hline
\end{tabular}
\end{table}
%More interestingly, with the aid of the DF strategy, the relaying link of the considered network has contribution to the achievable diversity gain.
%Compared with MABC, TDBC makes full use of the direct link and has lower operational complexity at the relay node, thus this work considers TDBC instead of MABC.
%All the channels follow independent but non-identically distributed Nakagami-$m$ fading.

The main contributions are summarized below.
\begin{itemize}
  \item Considering the impact of HIs, we obtain a closed-form expression for the system outage probability under independent but non-identically distributed (i.n.i.d.) Nakagami-$m$ fading channels.
      The derived system outage probability reveals an overall system ceiling (OSC) effect, i.e., when the target rate exceeds an OSC threshold that is determined by the HIs levels, the system falls into outage.
      %We refer such an effect and the corresponding value as the overall system ceiling (OSC) and the OSC threshold respectively.
      It is worth noting that, different from a SWIPT based AF TWRN that suffers from not only the OSC effect but also the relay cooperation ceiling (RCC) effect caused by HIs \cite{1234567}, the SWIPT based DF TWRN sees only the OSC effect.
  \item We derive the achievable diversity gain, which equals either zero or the sum of the shape parameter of the direct link and the smaller shape parameter of the terminal-to-relay links. This is quite different from the AF strategy, where the diversity gain from the relaying links is always zero.
      %Please note that, under the AF strategy, the diversity gain from the relaying link is always zero no matter what the transmission rate is.
  \item The analytical and simulation results reveal the following insights. The optimal PS ratio, which minimizes the system outage probability, increases as the channel quality of the relaying links improves. The relaying links under the DF strategy are more robust to HIs than that under the AF strategy. In the presence of HIs, the SWIPT based TWRN under the DF strategy outperforms that under the AF strategy in terms of the system outage probability.
\end{itemize}

%The remainder of this paper is organised as follows.  In Section II, we introduce the system and channel models, and then obtain the end-to-end siganl-to-noise-plus-distortion ratios. In Section III, we derive the system outage probability in a closed-form, and further analyze the diversity gain of the considered network. In Section IV, simulation results are provided. Finally, Section V concludes this paper.
In this paper, the important notations have been summarized in Table I.
\begin{figure}[!t]
  \centering
  \includegraphics[width=0.45\textwidth]{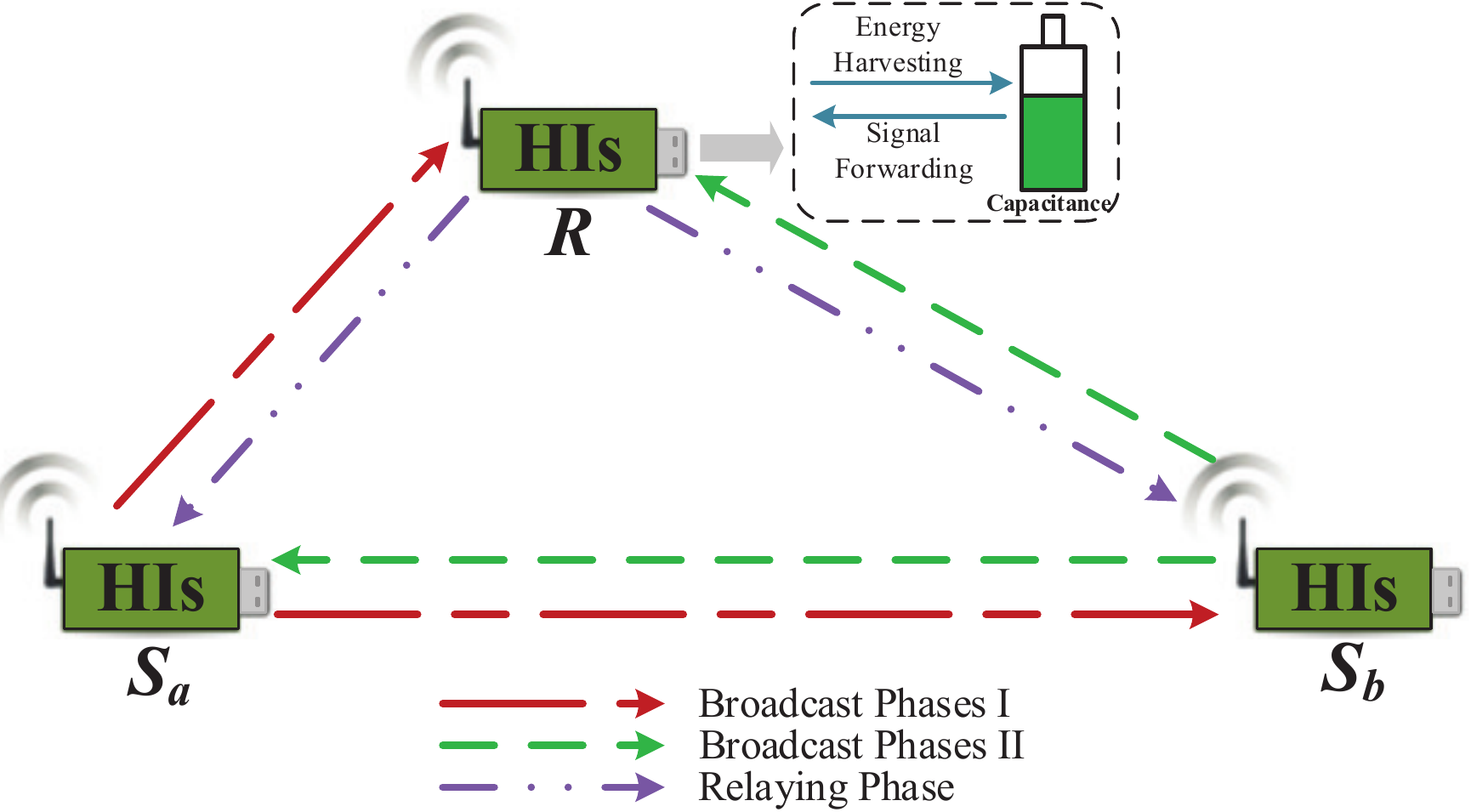}
  \caption{Model for the SWIPT based two-way DF relay network under a TDBC protocol.}
\end{figure}

%\begin{figure}[!t]
%  \centering
%  %\includegraphics[width=0.29\textwidth]{integralregion4} \\
%  \includegraphics[width=0.32\textwidth]{fig0} \\
%  \caption{Block diagram of direct link with HIs.}
%\end{figure}
\section{System Model and Working Flow}
\subsection{System Model and Channel Model}
We consider a SWIPT based two-way DF relay network under a TDBC protocol, where terminal $S_a$ communicates with terminal $S_b$ with the aid of an energy-limited relay $R$, as shown in Fig. 1. Herein, $R$ harvests energy from the incident RF signal following a PS scheme and a harvest-then-forward protocol.
%the energy-constrained relay $R$ harvests energy from the RF signal via a PS protocol and adopts the \lq\lq harvest-then-forward\rq\rq \;strategy to assist the information interchange between the two terminals $S_a$ and $S_b$. Each transmission block $T$ is divided into three phases: two broadcast (BC) phases ($2T/3$) and one relay (RL) phase ($T/3$).
We assume that all nodes are equipped with a single antenna and operate in the half-duplex mode.

In the TDBC protocol, each transmission block $T$ is split into three phases, viz., two broadcast (BC) phases (each of duration $T/3$) and one relaying (RL) phase ($T/3$).
During the first BC phase, $S_a$ transmits its signal $x_a$ to $R$ and $S_b$. In the second BC phase, $S_b$ broadcasts its signal $x_b$ to $R$ and $S_a$.
%In the first BC phase, terminal $S_a$ broadcasts its own signal $x_a$ to relay $R$ and terminal $S_b$. In the second BC phase, terminal $S_b$ transmits its signal $x_b$ to relay $R$ and terminal $S_a$.
After receiving the signal from $S_a$ or $S_b$, $R$ splits it into two parts, i.e., one part is  for energy harvesting (EH) and the other is for information decoding (ID). During the RL phase, $R$ combines the decoded signals $x_a$ and $x_b$ using the bit-wise XOR based encoding \cite{8851300}, and then broadcasts the combined signal to $S_a$ and $S_b$ using the total harvested energy.

We assume that all channels are quasi-static and reciprocal,
%All channels are assumed to be quasi-static and reciprocal,
%All the channels are reciprocal, quasi-static
%\footnote{The channel fading coefficient remains constant within a transmission block but independently may change in the next lock.}, reciprocal\footnote{The reciprocity means  $h_{ij}$=$h_{ji}$.}
and subject to i.n.i.d. Nakagami-$m$ fading.
%Specifically, the channel fading coefficient of the direct link between $S_i$ {\color{blue}to} $S_j$ is denoted by $h_{ij}\sim N\!aka(m_d , \Omega_d)$, with $i,j \in \left\{ {a,b} \right\},\;i \ne j$,
Specifically, $h_{ij}\sim {\rm{Naka}}(m_d , \Omega_d)$ and $h_{ir}\sim {\rm{Naka}}(m_i , \Omega_i)$ represent the channel fading coefficients of the $S_i \rightarrow S_j$ and $S_i \rightarrow R$ links, respectively, with $i,j \in \left\{ {a,b} \right\},\;i \ne j$.
Since $h_{ij}$ and $h_{ir}$ follow the Nakagami distribution, the corresponding channel gains $|{h_{ij}}{|^2}$ and $|{h_{ir}}{|^2}$ follow independent and non-identical gamma distributions, and their PDF and  CDF are expressed as
\begin{align}\label{2222}
  {f_V}\left( v \right) = \frac{1}{{\Gamma \left( {{m}} \right)\theta^{{m}}}}{v^{{m} - 1}}{e^{ - \frac{v}{{{\theta }}}}},
\end{align}
\begin{align}\label{2555}
{F_V}\left( v \right) = \frac{1}{{\Gamma \left( {{m}} \right)}}\gamma \left( {{m},\;\frac{v}{{{\theta}}}} \right),
\end{align}
where $m$ and $\theta$ denote the shape parameter$^3$\footnotetext[3]{In order to avoid complicated algebraic operations, integer shape parameters of Nakagami-$m$ fading are assumed in this work \cite{6630485,8851300,1234567}. Please note that relaxing this assumption can make the analysis more general, which will be studied in our future work.} and the scale parameter of random variable $V$, respectively, $V \in \left\{ {|{h_{ij}}{|^2},|{h_{ir}}{|^2}} \right\}$, $i,j \in \{{a,b} \},i \ne j$. Specially, when $V{\rm{ = }}|{h_{ij}}{|^2}$ or ${|{h_{ir}}{|^2}}$, $m = {m_d}$ or $ {{m_i}}$, and $\theta {\rm{ = }}{\Omega _d}/{m_d}$ or ${{\Omega _i}/{m_i}}$, respectively.

\subsection{Working Flow}
As mentioned above, $S_a$ and $S_b$ broadcast the signals $x_a$ and $x_b$ in the first second BC phases respectively.

For the direct link $\left({S_i}\mathop  \to {S_j}, i,j \in \left\{ {a,b} \right\},i \ne j\right)$, as illustrated in Fig. 2 at the top of the next page, the received signal at $S_j$ from $S_i$, in the presence of HIs, can be expressed as \cite{8851300}
\begin{align}\label{1}
{y_{ij}} = {h_{ij}}\left( {{x_i} + {\tau _{t,i}}} \right) + {\tau _{r,j}} + {n_{ij}},
\end{align}
where ${x_i}$ is the signal transmitted by $S_i$ with ${P_i} = {\mathbb {E}}\left\{ {{{\left| {{x_i}} \right|}^2}} \right\}$,  ${\tau _{t,i}} \sim \mathcal{CN}\left( {0,k_1^2{P_i}} \right)$ represents
%\footnote{$\mathcal{CN}\left( {a,b} \right)$ denotes the Gaussian distribution with mean $a$ and variance $b$.}
the distortion noise generated by the transmitter of $S_i$$^4$\footnotetext[4]{Actually, the total transmit power of $S_i$ equals $\left(1+k_1^2\right)P_i$, which is greater than the power of ${x_i}$. According to the HIs level region specified by the 3GPP LTE \cite{6630485}, viz., ${k_i} \in \left[ {0.08,0.175} \right]$, $i = \left\{ {1,2} \right\}$, $P_i\leq \frac {P_{S_i,\rm{max}}}{1+0.175^2}$ should hold to always meet the transmit power constraint $P_i+k_1^2{P_i} \leq P_{S_i,\rm{max}}$, where $P_{S_i,\rm{max}}$ denotes the maximum transmit power of $S_i$.}, ${\tau _{r,j}} \sim \mathcal{CN}\left( {0,k_2^2{P_i}|{h_{ij}}{|^2}} \right)$ represents the distortion noise generated by the receiver of $S_j$ and ${{n_{ij}}}\sim \mathcal{CN}\left( {0,\sigma^2} \right)$ is the additive white Gaussian noise (AWGN) at $S_j$. Note that $k_1$ and $k_2$ characterize the HIs levels of the transmitter and receiver, respectively, and are assumed to be the same for each transceiver \cite{6630485}. For simplicity, we assume that $P_a=P_b=P_o$ \cite{8613860}.

\begin{figure}[!t]
  \centering
  \includegraphics[width=0.27\textwidth]{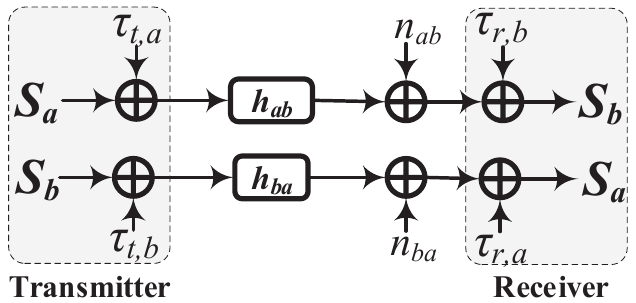}
  \caption{The signal flow diagram of the direct link under HIs.}
\end{figure}

\begin{figure}[!t]
  \centering
  \includegraphics[width=0.48\textwidth]{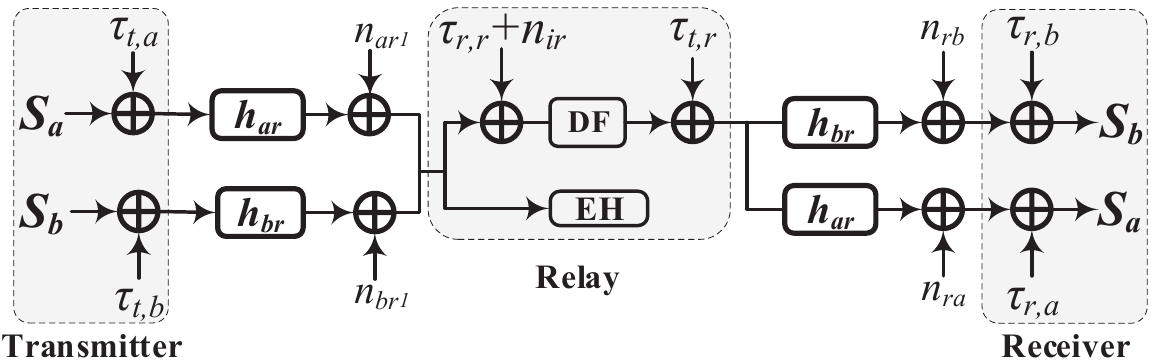}
  \caption{The signal flow diagram of the relaying link under HIs.}
\end{figure}

%For a given channel realization, the aggregate distortion noise at the receiver of $S_j$ has average power
%\begin{align}\label{2}
%{\mathbb {E}_{{\tau _{t,i}},{\tau _{r,j}}}}\left\{ {{{\left| {{h_{ij}}{\tau _{t,i}} + {\tau _{r,j}}} \right|}^2}} \right\} = \left( {k_1^2 + k_2^2} \right){P_i}{\left| {{h_{ij}}} \right|^2}.
%\end{align}
%
%According to \eqref{2}, \eqref{1} can be expressed, equivalently, as
%\begin{align}\label{3}
%{y_{ij}} = {h_{ij}}\left( {{x_i} + {\tau _{ij}}} \right) + {n_{ij}},
%\end{align}
%where ${\tau _{ij}} \sim \mathcal{CN}\left( {0,(k_1^2+k_2^2){P_i}} \right)$ denotes the aggregate distortion noise at the receiver of $S_j$.

Based on \eqref{1}, the signal-to-noise-plus-distortion ratio (SNDR) at $S_j$ from the direct link is written as
\begin{align}\label{4}
 {\gamma _{ij}} = \frac{{{\rho}|{h_{ij}}{|^2}}}{{\left( {k_1^2 + k_2^2} \right){\rho}|{h_{ij}}{|^2} + 1}},
\end{align}
where $\rho  = \frac{{{P_o}}}{{{\sigma ^2}}}$ denotes the input signal-to-noise ratio (SNR).

For the relaying link $\left({S_i}\mathop  \to \limits^R {S_j}, i,j \in \left\{ {a,b} \right\},i \ne j\right)$, as presented in Fig. 3, the received signal from $S_i$ at $R$, in the presence of HIs, can be written as
\begin{align}\label{5}
  {y_{ir}} = {h_{ir}} \left( {{x_i} + {\tau _{t,i}}} \right) + {n_{i{r_1}}},
\end{align}
where ${\tau _{t,i}} \sim \mathcal{CN}\left( {0,k_1^2{P_o}} \right)$ represents the hardware distortion noise generated by the transmitter of $S_i$ and ${n_{ir_1}} \sim \mathcal{CN}\left( {0,\sigma _{ir_1}^2} \right)$ is the antenna noise at $R$.

Using the PS protocol, the received signal ${y_{ir}}$ is split into two parts via a PS ratio $\beta$ ($0<\beta<1$): $\sqrt \beta {y_{ir}}$ for EH and $\sqrt {1 - \beta } {y_{ir}}$ for ID. Thus, all the harvested energy during the two BC phases can be expressed as \cite{8851300}
\begin{align}\label{6}
  {E_h} = \frac{T}{3}\eta \beta {P_o}\left( {|{h_{ar}}{|^2} + |{h_{br}}{|^2}} \right),
\end{align}
where $\eta\in(0,1)$ denotes the energy conversion efficiency. Based on \eqref{6}, the transmit power at $R$ can be calculated as
\begin{align}
{P_r} = \frac{{{E_h}}}{{T/3}} = \eta \beta {P_o}\left( {|{h_{ar}}{|^2} + |{h_{br}}{|^2}} \right).
\end{align}

Furthermore, at the relay $R$, the received signal from $S_i$  used for ID can be written as
\begin{align}\label{7}
y_{ir}^{{\rm{ID}}} = {h_{ir}}\sqrt {1 - \beta } \left( {{x_i} + {\tau _{t,i}}} \right) + {\tau _{r,r}} + {n_{ir}},
\end{align}
where ${\tau _{r,r}} \sim \mathcal{CN} \left( {0,k_2^2(1-\beta){P_o} {|{h_{ir}}{|^2}}} \right)$ represents the distortion noise$^5$\footnotetext[5]{Given that the vast majority of the receiver distortion noises are generated in the down-conversion process, we ignore the distortion noise induced by the receiving process at the antenna for analytical tractability \cite{7835665,1234567}.} generated by the receiver of $R$ and ${n_{ir}} \sim \mathcal{CN}\left( {0,\sigma^2} \right)$ is the AWGN at $R$.

Based on \eqref{7}, the SNDR for decoding $x_i$ at $R$ is given by
\begin{align}\label{8}
  {\gamma _{ir}} = \frac{{\left( {1 - \beta } \right)\rho {{\left| {{h_{ir}}} \right|}^2}}}{{\left( {k_1^2 + k_2^2} \right)\left( {1 - \beta } \right)\rho {{\left| {{h_{ir}}} \right|}^2} + 1}}.
\end{align}

If both $x_a$ and $x_b$ are successfully decoded during the two BC phases, $R$ performs bit-wise XOR based encoding to obtain the re-encoded signal $x_r$, i.e., ${x_r} = {x_a} \oplus {x_b}$ with ${P_r} = {\mathbb {E}}\left\{ {{{\left| {{x_r}} \right|}^2}} \right\}$ \cite{8851300}. Then $R$ broadcasts the signal $x_r$ to both terminals in the RL phase. Accordingly, the received signal at $S_i$ from $R$ can be written as
\begin{align}\label{38}
{y_{ri}} = {h_{ri}}\left( {{x_r} + {\tau _{t,r}}} \right){\rm{ + }}{\tau _{r,i}} + {n_{ri}},
\end{align}
where ${\tau _{t,r}} \sim \mathcal{CN}\left( {0,k_1^2P_r} \right)$ represents the distortion noise generated by the transmitter of $R$, ${\tau _{r,i}} \sim \mathcal{CN}\left( {0,k_2^2{P_r}{|{h_{ri}}{|^2}} } \right)$ represents the distortion noise generated by the receiver of $S_i$ and ${n_{ri}} \sim \mathcal{CN}\left( {0,\sigma^2} \right)$ is the AWGN at $S_i$.

\begin{figure*}[!t]
\normalsize
\setcounter{equation}{13}
\begin{align}\label{11}
{\mathbb{P}_{{{out}}}} = \underbrace {\Pr \left\{ {\left( {1 - \left( {k_1^2 + k_2^2} \right){\gamma _{th}}} \right)\rho Z < {\gamma _{th}}} \right\}}_{{\mathbb{P}_1}}\! \!\left( \!{1\! -\! \underbrace {\Pr \left\{ \begin{array}{l}
\left( {1 - \beta } \right)\left( {1\! -\! \left( {k_1^2\! +\! k_2^2} \right){\gamma _{th}}} \right)\rho X\! >\! {\gamma _{th}},\\
\left( {1 - \beta } \right)\left( {1\! -\! \left( {k_1^2\! +\! k_2^2} \right){\gamma _{th}}} \right)\rho Y\! >\! {\gamma _{th}},\\
\eta \beta \left( {1\! -\! \left( {k_1^2\! +\! k_2^2} \right){\gamma _{th}}} \right)\rho X\left( {X + Y} \right)\! >\! {\gamma _{th}},\\
\eta \beta \left( {1\! -\! \left( {k_1^2 \!+ \!k_2^2} \right){\gamma _{th}}} \right)\rho Y\left( {X + Y} \right)\! > \!{\gamma _{th}}
\end{array} \right\}}_{{\mathbb{P}_2}}}\! \right)
\end{align}
\setcounter{equation}{22}
\hrulefill
\setcounter{equation}{17}
\begin{align}\label{19}
&\mathbb{P}_2^1 =\frac{1}{{\Gamma \left( {{m_a}} \right)}}\Gamma \left( {{m_a},{\Delta _1}/{\theta _a}} \right) - \frac{1}{{\Gamma \left( {{m_a}} \right)\theta _a^{{m_a}}}}\sum\limits_{l = 0}^{{m_b} - 1} {\frac{1}{{l!}}} {\left( {\frac{1}{{{\theta _b}}}} \right)^l}{\left( {\frac{1}{{{\theta _a}}} + \frac{1}{{{\theta _b}}}} \right)^{ - \left( {l + {m_a}} \right)}}\Gamma \left( {l + {m_a},{\Delta _1}\left( {\frac{1}{{{\theta _a}}} + \frac{1}{{{\theta _b}}}} \right)} \right)\nonumber\\
&+ \frac{1}{{\Gamma \left( {{m_b}} \right)}}\Gamma \left( {{m_b},{\Delta _1}/{\theta _b}} \right) - \frac{1}{{\Gamma \left( {{m_b}} \right)\theta _b^{{m_b}}}}\sum\limits_{l = 0}^{{m_a} - 1} {\frac{1}{{l!}}} {\left( {\frac{1}{{{\theta _a}}}} \right)^l}{\left( {\frac{1}{{{\theta _a}}} + \frac{1}{{{\theta _b}}}} \right)^{ - \left( {l + {m_b}} \right)}}\Gamma \left( {l + {m_b},{\Delta _1}\left( {\frac{1}{{{\theta _a}}} + \frac{1}{{{\theta _b}}}} \right)} \right)\nonumber\\
&\!-\! \frac{1}{{\Gamma \left( \!{{m_b}} \!\right)}}\gamma \left(\! {{m_b},\!{\Delta _1}/{\theta _b}}\! \right)\!\left( \!{1\! -\! \frac{1}{{\Gamma \left( {{m_a}} \right)}}\gamma \left( {{m_a},{\Delta _1}/{\theta _a}} \right)} \right)\! -\! \frac{1}{{\Gamma \left(\! {{m_a}} \right)}}\gamma \left( {{m_a},{\Delta _1}/{\theta _a}} \right)\left( {1\! -\! \frac{1}{{\Gamma \left( {{m_b}} \right)}}\gamma \left( {{m_b},{\Delta _1}/{\theta _b}} \right)} \right)
\end{align}
\hrulefill
%\vspace*{-5pt}
\end{figure*}

According to \eqref{38}, the SNDR at $S_i$ from the relay link can be expressed as
\begin{align}\label{9}
\setcounter{equation}{10}
{\gamma _{ri}} = \frac{{\eta \beta \rho {{\left| {{h_{ir}}} \right|}^2}\left( \!{{{\left| {{h_{ar}}} \right|}^2}\! +\! {{\left| {{h_{br}}} \right|}^2}} \!\right)}}{{\left( {k_1^2\! +\! k_2^2} \right)\eta \beta \rho {{\left| {{h_{ir}}} \right|}^2}\left( {{{\left| {{h_{ar}}} \right|}^2}\! + \!{{\left| {{h_{br}}} \right|}^2}} \right) \!+\! 1}}.
\end{align}

Following the selection combining scheme, the end-to-end SNDR at $S_i$ can be written as
\begin{align}
{\gamma _i} = \max \left\{ {{\gamma _{ji}},\min \left\{ {{\gamma _{jr}},{\gamma _{ri}}} \right\}} \right\},
\end{align}
where $i,j \in \left\{ {a,b} \right\},i \ne j$.
%Note that ${\gamma _{ij}}={\gamma _{ji}}$ due to the reciprocity of channels.

\section{System Outage Performance Analysis}
In this section, we first derive the system outage probability in closed-form, then identify the OSC effect, and study the achievable diversity gain of our studied network.
\subsection{System Outage Probability}
The outage event occurs if the data rate $\mathbf{R}_a$ from $S_a$ to $S_b$ and/or the data rate $\mathbf{R}_b$ from $S_b$ to $S_a$ falls below a threshold $\mathbf{R}_{th}$ \cite{8633928}. Hence, the system outage probability$^6$, $\mathbb{P}_{out}$\footnotetext[6]{According to the definition of outage event, $\mathbb{P}_{out}$ can also be expressed as ${{\rm{\mathbb{P}}}_{out}} = \underbrace {\Pr \left\{ {{\gamma _a} < {\gamma _{th}}} \right\}}_{{\Lambda _1}} + \underbrace {\Pr \left\{ {{\gamma _b} < {\gamma _{th}}} \right\}}_{{\Lambda _2}}- \underbrace {\Pr \left\{ {{\gamma _a} < {\gamma _{th}},{\gamma _b} < {\gamma _{th}}} \right\}}_{{\Lambda _3}}$, where ${{\Lambda _1}}$ and ${{\Lambda _2}}$ denote the T2T outage probabilities at $S_a$ and $S_b$ respectively, and ${{\Lambda _3}}$ denotes the probability that both $S_a$ and $S_b$ are in outage. Since there is a high correlation between two T2T links, viz., ${S_a} \to {S_b}$ and ${S_b} \to {S_a}$ links, ${\Lambda _3} \ne \Pr \left\{ {{\gamma _a} < {\gamma _{th}}} \right\}\Pr \left\{ {{\gamma _b} < {\gamma _{th}}} \right\}$, which means that the system outage probability cannot be directly derived from T2T outage probabilities and the theoretical analysis for the system outage probability is more challenging.}, can be written as
\begin{align}\label{10}
\setcounter{equation}{12}
&{\mathbb{P}_{out}} = \Pr \left\{ {\min \left\{ {{{\mathbf{R}}_a},{{\mathbf{R}}_b}} \right\} < {{\mathbf{R}}_{th}}} \right\}\nonumber\\
&= \Pr \left\{ {\min \left\{ {{\gamma _a},{\gamma _b}} \right\} < {\gamma _{th}}} \right\}= \Pr \left\{ {{\gamma _{ab}} < {\gamma _{th}}} \right\}\nonumber\\
&\!\times\! \left( \!{1\! -\! \Pr \left\{ \!{{\gamma _{ar}} \!>\! {\gamma _{th}},{\gamma _{br}}\! >\! {\gamma _{th}},{\gamma _{ra}}\! > \!{\gamma _{th}},{\gamma _{rb}}\! >\! {\gamma _{th}}} \!\right\}}\! \right),
\end{align}
where  ${\gamma _{th}} = {2^{{{3{{\mathbf{R}}_{{{th}}}}} \mathord{\left/
 {\vphantom {{3{{\mathbb{R}}_{{{th}}}}} T}} \right.
 \kern-\nulldelimiterspace} T}}} - 1$ denotes the SNDR threshold and ${\mathbf{R}}_i=\frac{T}{3}{\log _2}\left( {1 + {\gamma _i}} \right)$, $i \in \{ a,b\}$.

Letting $|{h_{ar}}{|^2} = X$, $|{h_{br}}{|^2} = Y$ and $|{h_{ab}}{|^2} = Z$, and substituting \eqref{4}, \eqref{8} and \eqref{9} into \eqref{10}, ${\mathbb{P}_{out}}$ can be rewritten as \eqref{11}.

\begin{figure*}[!t]
\normalsize
\setcounter{equation}{18}
\begin{align}\label{21}
&\mathbb{P}_2^2 = \frac{{\pi \left( {\Phi  - \sqrt {{\Delta _2}/2} } \right)}}{{2N\Gamma \left( {{m_a}} \right)\theta _a^{{m_a}}}}\sum\limits_{l = 0}^{{m_b} - 1} {\sum\limits_{n = 1}^N {\sqrt {1 - v_n^2} \frac{1}{{l!}}{{\left( {\frac{1}{{{\theta _b}}}} \right)}^l}} } {\left( {k_n^{\rq} + \sqrt {{\Delta _2}/2} } \right)^{{m_a} - 1}}{Q^l}\left( {k_n^{\rq} + \sqrt {{\Delta _2}/2} } \right)\nonumber\\
&\times {e^{ - \frac{{k_n^{\rq} + \sqrt {{\Delta _2}/2} }}{{{\theta _a}}} - \frac{{Q\left( {k_n^{\rq} + \sqrt {{\Delta _2}/2} } \right)}}{{{\theta _b}}}}} - \frac{1}{{\Gamma \left( {{m_a}} \right)\theta _a^{{m_a}}}}\sum\limits_{l = 0}^{{m_b} - 1} {\frac{1}{{l!}}} {\left( {\frac{1}{{{\theta _b}}}} \right)^l}{\left( {\frac{{{\theta _a}{\theta _b}}}{{{\theta _a} + {\theta _b}}}} \right)^{{m_a} + l}}\bigg(\gamma \left( {{m_a} + l,\Phi \left( {\frac{1}{{{\theta _a}}} + \frac{1}{{{\theta _b}}}} \right)} \right)\nonumber\\
& - \gamma \left( {{m_a} + l,\sqrt {{\Delta _2}/2} \left( {\frac{1}{{{\theta _a}}} + \frac{1}{{{\theta _b}}}} \right)} \right)\bigg) + \frac{{\pi \left( {\Phi  - \sqrt {{\Delta _2}/2} } \right)}}{{2N\Gamma \left( {{m_b}} \right)\theta _b^{{m_b}}}}\sum\limits_{l = 0}^{{m_a} - 1} {\sum\limits_{n = 1}^N {\sqrt {1 - v_n^2} \frac{1}{{l!}}{{\left( {\frac{1}{{{\theta _a}}}} \right)}^l}} } {\left( {k_n^{\rm{'}} + \sqrt {{\Delta _2}/2} } \right)^{{m_b} - 1}}\nonumber\\
&\times {Q^l}\left( {k_n^{{\rq}} + \sqrt {{\Delta _2}/2} } \right){e^{ - \frac{{k_n^{{\rq}} + \sqrt {{\Delta _2}/2} }}{{{\theta _b}}} - \frac{{Q\left( {k_n^{{\rq}} + \sqrt {{\Delta _2}/2} } \right)}}{{{\theta _a}}}}} - \frac{1}{{\Gamma \left( {{m_b}} \right)\theta _b^{{m_b}}}}\sum\limits_{l = 0}^{{m_a} - 1} {\frac{1}{{l!}}} {\left( {\frac{1}{{{\theta _a}}}} \right)^l}{\left( {\frac{{{\theta _a}{\theta _b}}}{{{\theta _a} + {\theta _b}}}} \right)^{{m_b} + l}}\nonumber\\
&\times \left( {\gamma \left( {{m_b} + l,\Phi \left( {\frac{1}{{{\theta _a}}} + \frac{1}{{{\theta _b}}}} \right)} \right) - \gamma \left( {{m_b} + l,\sqrt {{\Delta _2}/2} \left( {\frac{1}{{{\theta _a}}} + \frac{1}{{{\theta _b}}}} \right)} \right)} \right)\nonumber\\
&+ \frac{1}{{\Gamma \left( {{m_a}} \right)}}\Gamma \left( {{m_a},\Phi /{\theta _a}} \right) - \frac{1}{{\Gamma \left( {{m_a}} \right)\theta _a^{{m_a}}}}\sum\limits_{l = 0}^{{m_b} - 1} {\frac{1}{{l!}}} {\left( {\frac{1}{{{\theta _b}}}} \right)^l}{\left( {\frac{1}{{{\theta _a}}} + \frac{1}{{{\theta _b}}}} \right)^{ - \left( {l + {m_a}} \right)}}\Gamma \left( {l + {m_a},\Phi \left( {\frac{1}{{{\theta _a}}} + \frac{1}{{{\theta _b}}}} \right)} \right)\nonumber\\
&+ \frac{1}{{\Gamma \left( {{m_b}} \right)}}\Gamma \left( {{m_b},\Phi /{\theta _b}} \right) - \frac{1}{{\Gamma \left( {{m_b}} \right)\theta _b^{{m_b}}}}\sum\limits_{l = 0}^{{m_a} - 1} {\frac{1}{{l!}}} {\left( {\frac{1}{{{\theta _a}}}} \right)^l}{\left( {\frac{1}{{{\theta _a}}} + \frac{1}{{{\theta _b}}}} \right)^{ - \left( {l + {m_b}} \right)}}\Gamma \left( {l + {m_b},\Phi \left( {\frac{1}{{{\theta _a}}} + \frac{1}{{{\theta _b}}}} \right)} \right)\nonumber\\
&\!- \!\frac{1}{{\Gamma \left( {{m_b}} \right)}}\gamma \left( {{m_b},{\Delta _1}/{\theta _b}} \right)\!\left( {1\! -\! \frac{1}{{\Gamma \left( {{m_a}} \right)}}\gamma \left( {{m_a},\!\Phi /{\theta _a}} \right)} \right)\! -\! \frac{1}{{\Gamma \left( {{m_a}} \right)}}\gamma \left( {{m_a},{\Delta _1}/{\theta _a}} \right)\left( {1\! -\! \frac{1}{{\Gamma \left( {{m_b}} \right)}}\gamma \left( {{m_b},\Phi /{\theta _b}} \right)} \right)
\end{align}
\setcounter{equation}{22}
\hrulefill
%\vspace*{-5pt}
\end{figure*}

Next, we derive $\mathbb{P}_1$ and $\mathbb{P}_2$ to obtain the closed-form expression for $\mathbb{P}_{out}$ respectively. The first term of \eqref{11}, $\mathbb{P}_1$, can be calculated as
\begin{align}\label{12}
\setcounter{equation}{14}
  {\mathbb{P}_1} = \left\{ \begin{array}{l}
{\mathbb{P}_{1}^1}\;,\;\;{\gamma _{th}} < \frac{1}{{k_1^2 + k_2^2}},\\
\;1\;\;,\;\;{\gamma _{th}} \ge \frac{1}{{k_1^2 + k_2^2}},
\end{array} \right.
\end{align}
where ${\mathbb{P}_{1}^1}$ is given by
\begin{align}\label{13}
  {\mathbb{P}_{1}^1} =& 1 - {e^{ - \frac{{{\gamma _{th}}}}{{{\theta _d}\rho \left( {1 - {\gamma _{th}}\left( {k_1^2 + k_2^2} \right)} \right)}}}}\nonumber\\
  &\times \sum\limits_{l = 0}^{{m_d}-1} {\frac{1}{{l!}}} {\left( {\frac{{{\gamma _{th}}}}{{{\theta _d}\rho \left( {1 - {\gamma _{th}}\left( {k_1^2 + k_2^2} \right)} \right)}}} \right)^l}.
\end{align}

\emph{Proof.} Please refer to Appendix A. \hfill {$\blacksquare $}

The second term of \eqref{11}, $\mathbb{P}_2$, can be calculated as
\begin{align}\label{14}
\setcounter{equation}{16}
{\mathbb{P}_2} = \left\{ \begin{array}{l}
{\mathbb{P}_{2}^*}\;,\;\;{\gamma _{th}} < \frac{1}{{k_1^2 + k_2^2}},\\
\;0\;\;,\;\;{\gamma _{th}} \ge \frac{1}{{k_1^2 + k_2^2}},
\end{array} \right.
\end{align}
where ${\mathbb{P}_{2}^*}$ equals either ${\mathbb{P}_2^1}$ or ${\mathbb{P}_2^2}$, as presented respectively in \eqref{19} at the top of the previous page and \eqref{21}. Please note that, ${\mathbb{P}_2^1}$ corresponds to the case that ${\Delta _{\rm{1}}} \ge \sqrt {{\Delta _{\rm{2}}}{\rm{/2}}}$ is satisfied (as discussed in Appendix B), while ${\mathbb{P}_2^2}$ corresponds to the case that ${\Delta _{\rm{1}}} < \sqrt {{\Delta _{\rm{2}}}{\rm{/2}}}$ holds, where ${\Delta_1} = \frac{{{\gamma _{th}}}}{{\left( {1 - \beta } \right)\left( {1 - \left( {k_1^2 + k_2^2} \right){\gamma _{th}}} \right)\rho }}$ and ${\Delta_2} = \frac{{{\gamma _{th}}}}{{\eta \beta \left( {1 - \left( {k_1^2 + k_2^2} \right){\gamma _{th}}} \right)\rho }}$.

\emph{Proof.} Please refer to Appendix B. \hfill {$\blacksquare $} 

Substituting \eqref{12} and \eqref{14} into \eqref{11}, the system outage probability of the considered network can be expressed as
\begin{align}\label{16}
\setcounter{equation}{19}
{\mathbb{P}_{out}} = \left\{ \begin{array}{l}
{\mathbb{P}_{1}^1}\left(1-{\mathbb{P}_{2}^*}\right)\;,\;\;{\gamma _{th}} < \frac{1}{{k_1^2 + k_2^2}},\\
\;\;\;\;\;\;\;\;\;1\;\;\;\;\;\;\;\;,\;\;{\gamma _{th}} \ge \frac{1}{{k_1^2 + k_2^2}}.
\end{array} \right.
\end{align}

\begin{remark}\label{37}
As illustrated in \eqref{16}, HIs deteriorate the system outage performance by imposing a constraint on $\gamma_{th}$. In particular, when ${\gamma _{th}} < \frac{1}{{k_1^2 + k_2^2}}$, both the direct and relaying links participate in the information exchange between two terminals. However, when ${\gamma _{th}} \ge \frac{1}{{k_1^2 + k_2^2}}$, the overall system ceases no matter what the input SNR is. This effect is referred to as OSC, and $\frac{1}{{k_1^2 + k_2^2}}$ denotes the corresponding OSC threshold. This ceiling effect is due to the fact that the instantaneous SNDRs at $S_a$, $S_b$ and $R$, given in \eqref{4}, \eqref{8} and \eqref{9}, are upper bounded by ${\gamma _{ij}}$, ${\gamma _{ir}}$, ${\gamma _{rj}} < \frac{1}{{k_1^2 + k_2^2}}$, respectively, $i,j \in \{ a,b\}$ and $i \ne j$. In addition, it is obvious that the maximum achievable SNDR threshold $\gamma^{max}_{th}\approx\frac{1}{{k_1^2 + k_2^2}}$ decreases with the increase of the HIs levels.
%Specifically, when $\gamma_{th}$ is below $\frac{1}{{k_1^2 + k_2^2}}$, both the direct and relaying links cooperate with the information exchange between two terminals. When $\gamma_{th}$ exceeds $\frac{1}{{k_1^2 + k_2^2}}$, the system is in outage no matter what the input SNR is. This effect is referred to as OSC. Obviously, the threshold of $\gamma_{th}$ corresponding to OSC is determined by HIs levels.
\end{remark}

\subsection{Diversity Gain}
Referring to \cite{8879698}, the achievable diversity gain of the considered network can be given as
\begin{align}\label{35}
\setcounter{equation}{20}
  d =  - \mathop {\lim }\limits_{\rho  \to \infty } \frac{{\log \left( {{\mathbb{P}_{out}}} \right)}}{{\log \left( \rho  \right)}}.
\end{align}

Substituting \eqref{16} into \eqref{35}, the diversity gain can be calculated as
\begin{align}\label{131}
  d = \left\{ \begin{array}{l}
m_d+\min \left( {{m_a},{m_b}} \right)\;\;,\;{\gamma _{th}} < \frac{1}{{k_1^2 + k_2^2}},\\
\;\;\;\;\;\;\;\;\;\;\;\;\;\;0\;\;\;\;\;\;\;\;\;\;\;\;\;\;\;\;\;,\;{\gamma _{th}} \ge \frac{1}{{k_1^2 + k_2^2}}.
\end{array} \right.
\end{align}

\emph{Proof.} Please refer to Appendix C. \hfill {$\blacksquare $}  

\begin{remark}\label{37}
%As presented in \eqref{131}, the HIs causes the detrimental impact on the achievable diversity gain by posing constraint on ${\gamma _{th}}$.
Eq. \eqref{131} reveals the following facts. When ${\gamma _{th}}$ exceeds the OSC threshold, vie., ${\gamma _{th}} \geq \frac{1}{{k_1^2 + k_2^2}}$, the diversity gain of the overall system is equal to zero. When ${\gamma _{th}} < \frac{1}{{k_1^2 + k_2^2}}$, the diversity gain is the sum of the shape parameter of the direct link and the smaller shape parameter of the terminal-to-relay links.\;In other words, both the direct and relaying links jointly enhance the diversity gain. However, for the AF strategy, HIs makes the diversity gain from the relaying links zero \cite{1234567}. The reason is as follows. For the AF strategy, one observation from [20, eq. (10)] is that the SNDR of the relaying link at high $\rho$ is a function of channel fading coefficients,  which cannot ensure that  the SNDR is larger than its threshold with probability one even though $\rho \to \infty $. Accordingly, the achievable
 diversity gain of AF relaying links equals zero. Such a result is mainly due to the fact that the AF strategy amplifies the signals and distortion noises simultaneously. However, in the DF relaying, the SNDRs of both the relay and terminals at high $\rho$ keep constant, i.e., ${\gamma _{ir}}={\gamma _{rj}}=\frac{1}{{k_1^2 + k_2^2}}$. When ${\gamma _{th}} < \frac{1}{{k_1^2 + k_2^2}}$, the outage probability for the relaying links approaches zero with the increase of $\rho$, which allows the relaying links to contribute to the diversity gain.
%However, in the DF relaying, the SNDRs of the both relay $R$ and terminals at high $\rho$ keep constant, which allows the relaying link to contribute to the diversity gain when ${\gamma _{th}} < \frac{1}{{k_1^2 + k_2^2}}$.
\end{remark}

\section{Simulation Results}
In this section, numerical results are provided to validate the correctness of the above theoretical analyses.
%simulation results are provided to validate the above theoretical analysis and study the impact of various parameters on the system outage performance.
Hereinafter, unless otherwise specified, the simulation parameters are set as Table 2 \cite{8851300,1234567,6952965}.
%: $\eta  = 0.6$, $\beta=0.9$, $T = 1$ \!sec and ${k_1} = {k_2} = {k_{ave}}$. Moreover, the distances of $S_a-R$, $S_b-R$ and $S_a-S_b$ links are set as $d_{ar}=5$ m, $d_{br}=5$ m and $d_{ab}=10$ m, respectively. The average powers $\Omega_a$, $\Omega_b$ and $\Omega_{d}$ are expressed as ${\Omega _a} = d_{ar}^{ - {\alpha _1}}$, ${\Omega _b} = d_{br}^{ - {\alpha _1}}$ and ${\Omega _d} = d_{ab}^{ - {\alpha _2}}$, where $\alpha_1=2.7$ and $\alpha_2=3$ denote the path loss exponents of the relaying links and the direct link, respectively.

\begin{table}[t]
\caption{Simulation parameters}
\centering
\begin{tabular}{|c|c|} %l(left)居左显示 r(right)居右显示 c居中显示
\hline
Parameter&Value\\
\hline
energy conversion efficiency, $\eta$&0.6\\
\hline
PS ratio, $\beta$&0.9\\
\hline
transmission block, $T$&1 s\\
\hline
HIs levels, $k_1$ and $k_2$&$k_{ave}$\\
\hline
the distances of $S_a-R$ link, $d_{ar}$&5 m\\
\hline
the distances of $S_b-R$ link, $d_{br}$&5 m\\
\hline
the distances of $S_a-S_b$ link, $d_{ab}$&10 m\\
\hline
the path loss exponents of the&2.7\\
 relaying links, $\alpha_1$& \\
\hline
the path loss exponents of the&3\\
  direct link, $\alpha_2$& \\
\hline
the average power, $\Omega_a$&$d_{ar}^{ - {\alpha _1}}$\\
\hline
the average power, $\Omega_b$&$d_{br}^{ - {\alpha _1}}$\\
\hline
the average power, ${\Omega _d}$&$d_{ab}^{ - {\alpha _2}}$\\
\hline
\textcolor{black}{the noise power $\sigma^2$}& \textcolor{black}{-50 dBm}\\
\hline
the channel bandwidth & 1 MHz\\
\hline
\end{tabular}
\end{table}

Fig. 4 verifies the derived system outage probability in all two cases of the \eqref{16}. Based on the used simulation parameters, we determine the specific value for the OSC threshold, viz., $\frac{1}{{k_1^2 + k_2^2}} = 50$. In addition, we also consider four different transmission rates i.e., $\mathbb{R}_{th}=1$, $1.5$, $2$ and $2.5$~bit/Hz. On the basis of the definition in \eqref{10}, the corresponding SNDR thresholds are calculated as $7$, $22$, $63$, and $180$ respectively.
%corresponding to four SNDR thresholds, viz. $7$ dB, $22$ dB, $63$ dB, and $180$ dB.
%Four different data rates are used, i.e., $1$~bit/Hz, $1.5$~bit/Hz, $2$~bit/Hz and $2.5$~bit/Hz, corresponding to four SNDR thresholds, viz. $7$ dB, $22$ dB, $63$ dB, and $180$ dB. We also determine the threshold constraints cased by the HIs, viz., $\frac{1}{{k_1^2 + k_2^2}} = 50$ dB.
According to \eqref{16}, when the SNDR threshold $\gamma_{th}=7 \& 22$ or $63 \& 180$, the corresponding system outage probability is expressed as ${\mathbb{P}_{1}^1}\left(1-{\mathbb{P}_{2}^*}\right)$ or $1$.
%Based on \eqref{16}, the corresponding system outage probability for above transmission rates can be expressed as
%%the system outage probability is computed as
%\begin{align}\notag
%{\mathbb{P}_{out}} = \left\{ \begin{array}{l}
%{\mathbb{P}_{1}^1}\left(1-{\mathbb{P}_{2}^1}\right)\;,\;\;{\gamma _{th}} = 7\;{\rm{dB}}\;{\rm{or}}\;22\;{\rm{dB}},\\
%\;\;\;\;\;\;\;\;1\;\;\;\;\;\;\;\;\;,\;\;{\gamma _{th}} = 63\;{\rm{dB}}\;{\rm{or}}\;180\;{\rm{dB}}.
%\end{array} \right.
%\end{align}
As presented in Fig. 4, the \lq $\lozenge$, $\Box$, $\triangledown$, $\circ$\rq \;marked curves (analytical results) match precisely with the \lq -\rq\;marked curves (simulation results) across the entire region, which validates the correctness of the derived system outage probability in Section 3.1.
%As shown in Fig. 1, the analytical results match with simulation results well. This verifies the accuracy of the derived closed-form expression.

\begin{figure}[!t]
  \centering
  \includegraphics[width=0.45\textwidth]{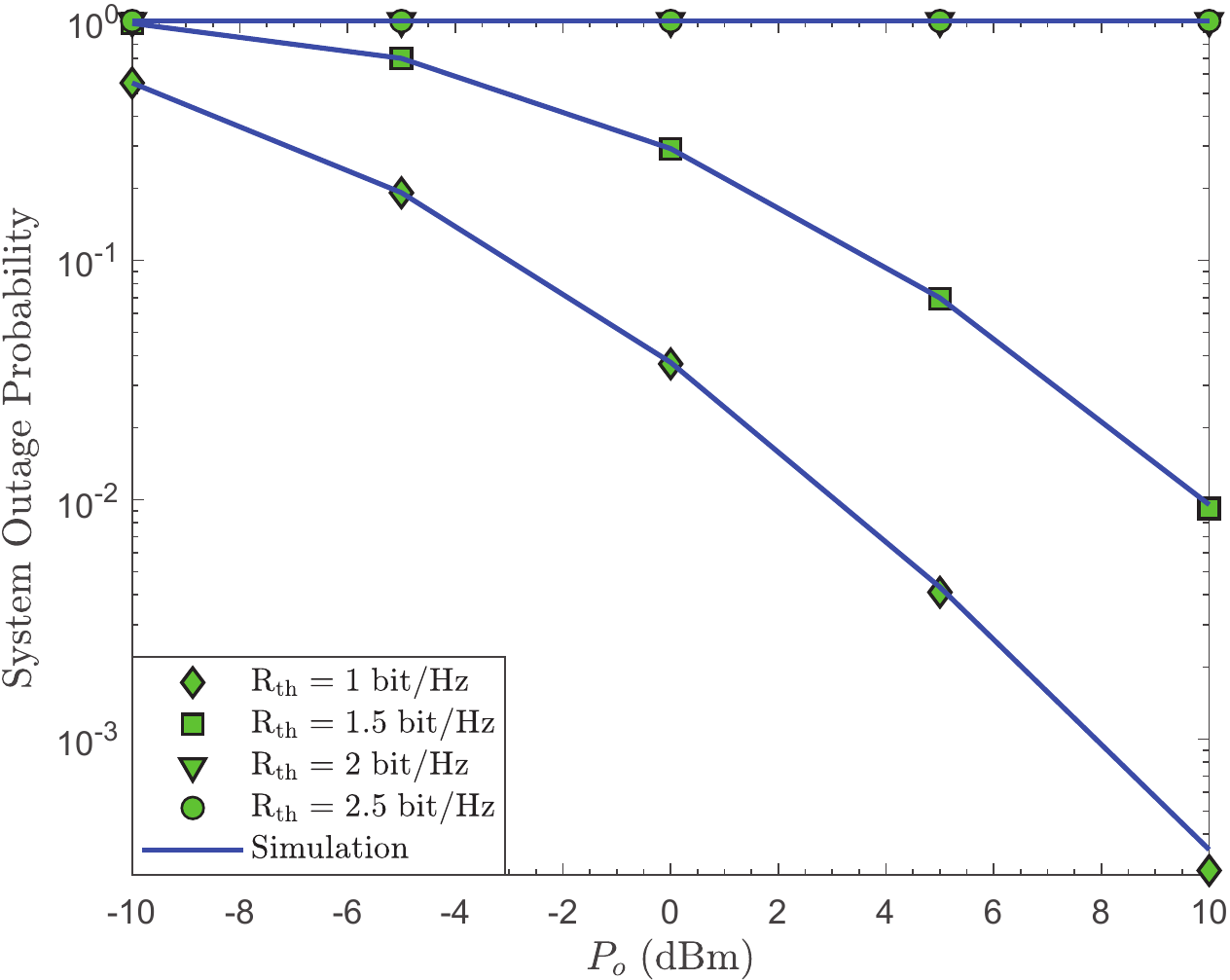} \\
  \caption{The derived system outage probability ${\mathbb{P}_{out}}$ vs. the transmit power of each terminal $P_o$, in comparison with simulation results, where $k_{ave}=0.1$ and $\{m_a, m_b, m_d\}=\{2, 2, 1\}$.}
\end{figure}

\begin{figure}[!t]
  \centering
  \includegraphics[width=0.45\textwidth]{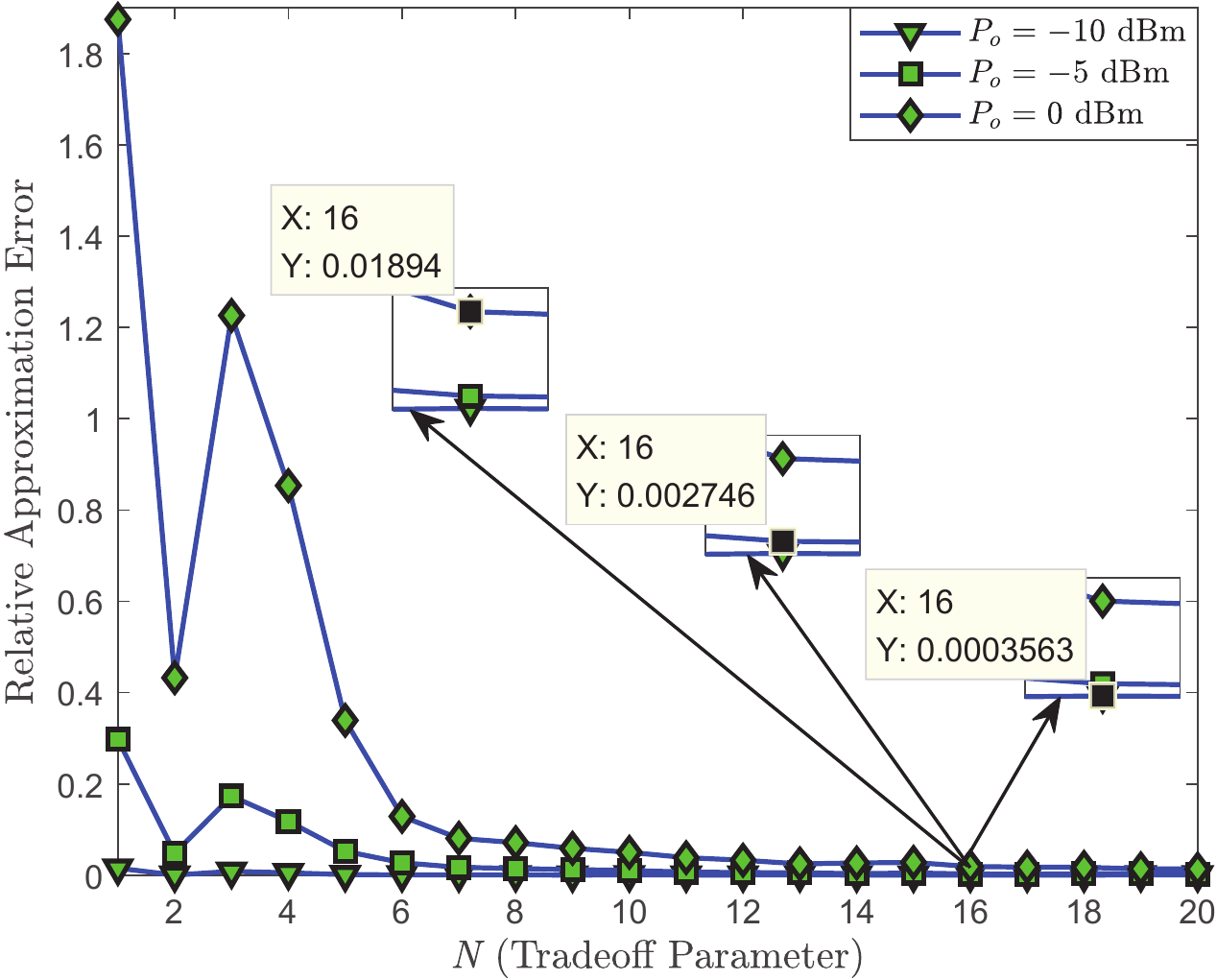} \\
  \caption{The relative approximation error vs. the trade-off parameter $N$, where $k_{ave}=0.1$, $\mathbb{R}_{th}=1$ bit/Hz and $\{m_a, m_b, m_d\}=\{2, 2, 1\}$.}
\end{figure}

\begin{figure}[!t]
  \centering
  \includegraphics[width=0.45\textwidth]{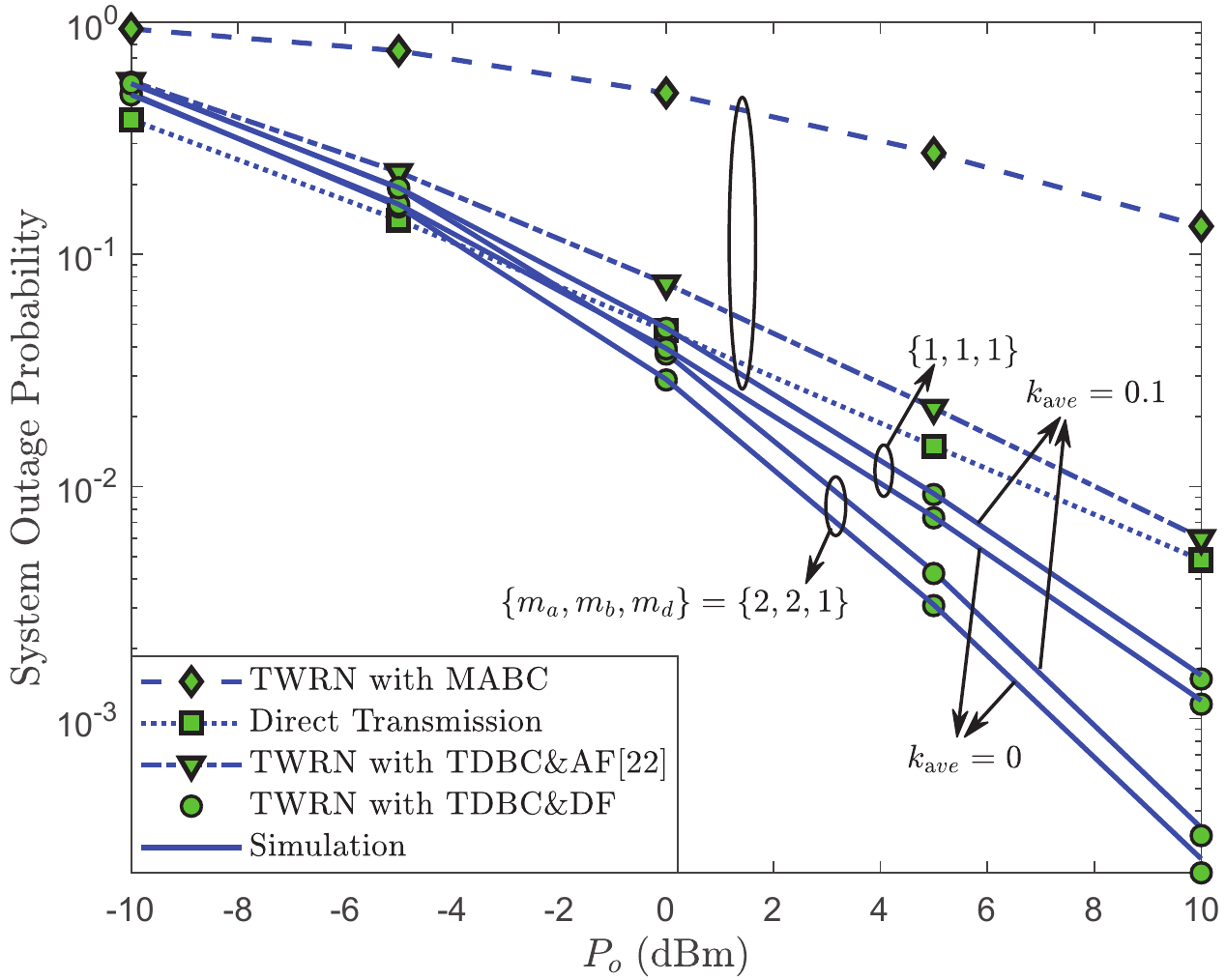} \\
  \caption{The system outage probability ${\mathbb{P}_{out}}$ vs. the transmit power of each terminal $P_o$, where $\mathbb{R}_{th}=1$ bit/Hz.}
\end{figure}

%Fig. 4 shows the relative approximation error against tradeoff parameter $N$ with different input SNRs to illustrate the performance of Gaussian-Chebyshev quadrature approximation approach.
In Appendix B, the Gaussian-Chebyshev quadrature approach is adopted to obtain the approximate system outage probability ${\mathbb{P}_2^2}$ in \eqref{16} when ${\Delta _{\rm{1}}} < \sqrt {{\Delta _{\rm{2}}}{\rm{/2}}}$.
To illustrate the performance of the Gaussian-Chebyshev quadrature approximation approach, Fig. 5 shows the relative approximation error against the tradeoff parameter $N$ for different transmit power of each terminal.
Specially, the relative approximation error $\delta$ equals the ratio of the absolute value of difference between the simulation and analytical results to the simulation result,
%, where the relative approximation error, $\delta$, is defined as \cite{8633928}
%\begin{align}\notag
%\delta  = \left| {\frac{{{\rm{simulation\;result - analytical\;result}}}}{{{\rm{simulation\;result}}}}} \right|,
%\end{align}
in which the simulation and analytical results are obtained by compute simulation and \eqref{16} respectively. From Fig. 5, one can observe that $\delta$ gradually approaches zero as $N$ increases.\;For instance, when $P_o\;=\;-5$ dBm and $N=16$, the corresponding $\delta$ is $0.002746$, which demonstrates that the Gaussian-Chebyshev quadrature with few terms can evaluate the system outage performance precisely.
%the approximate result is sufficiently accurate. Hence, the derived expression based on Gaussian-Chebyshev quadrature approximation approach can be used to evaluate the system outage performance efficiently.

Fig. 6 presents the system outage probability versus transmit power of each terminal $P_o$ for four transmission protocols, viz., the TDBC with DF strategy (viz., the considered network), the TDBC with AF strategy \cite{1234567}, the direct transmission, and the MABC.
%\footnote{In the direct transmission protocol, each transmission block is divided into two equal phases. In the first phase, terminal $S_a$ transmits its signal to terminal $S_b$, while terminal $S_b$ transmit its signal to terminal $S_a$ in the second phase.}.
%compares the system outage performance among three transmission protocols, namely the TDBC, the MABC, and the direct transmission, under different HIs levels and shape parameters.
%Specially, in the direct transmission protocol, each transmission block is divided into two phases. The terminal $S_a$ broadcasts its own signal to the terminal $S_b$ in the first phase, whereas The terminal $S_b$ broadcasts its own signal to the terminal $S_a$ in the second phase.
Herein, we ensure that the used target transmission rate satisfies ${\gamma _{th}} < \frac{1}{{k_1^2 + k_2^2}}$.
%ensure that the used SNDRs threshold is below that of OSC. Moreover, for a fair comparison, the four transmission protocols assume the same transmission rate threshold and the same energy consumption at each terminal.
%Moreover, in order to fair comparison, we also assume that the four transmission protocols consume the same energy and transmission rate at each terminal.
As given in Fig. 6, the considered network achieves the higher system outage performance than both the MABC and the direct transmission.
%This is due to the MABC and direct transmission only live on the relaying link and the direct link respectively, but the considered TDBC fully use both the direct and relaying links.
One can also note that the system outage performance of the considered network outperforms that of the TDBC with AF strategy. This suggests that under the presence of HIs, the DF strategy is superior to AF strategy in terms of system outage performance. Additionally, it can be observed that, for the given shape parameters, the considered network exists a performance gap between the curves corresponding to ideal hardware ($k_{ave}=0$) and HIs ($k_{ave}=0.1$), which concludes that the HIs degrade the system outage performance to some extent. Finally, as expected, better channel quality of the relaying links can significantly improve the system outage performance for the considered network under a fixed HIs level.

%Additionally, as expected, better channel quality of the relaying links can significantly improve the system outage performance for the considered network under a fixed HIs levels. Finally, it can be observed that, for the given shape parameters, the considered network exist a performance gap between the curves corresponding to ideal hardware ($k_{ave}=0$) and HIs ($k_{ave}=0.1$), which concludes that the HIs deteriorates the system performance.

%Clearly, we can see that the  considered TDBC achieves the lowest system outage probability while the system outage probability of the MABC is the highest. This is because TDBC can fully use both the direct and relaying links, however, the MABC and direct transmission only rely on the relaying link and the direct link, respectively.
%%Note that the system outage probability for the TWRN with MABC tends to a constant in high SNR regions.
%Also, we investigate the effects of shape parameters at relaying link and HIs levels on the system outage probability of the considered TWRN with TDBC. In particular, for the given HIs levels, the system outage probability decreases with the increase of shape parameters $m_a$ and $m_b$. This is because the relaying link conditions get better as the increase of $m_a$ and $m_b$, which in turn improve the system outage probability. In addition,  for the given shape parameters, the system outage probability with HIs ($k_{ave}=0.1$) is much higher than that with ideal hardware ($k_{ave}=0$). This indicates that the HIs seriously deteriorates the system outage performance.
\begin{figure}[!t]
  \centering
  \includegraphics[width=0.45\textwidth]{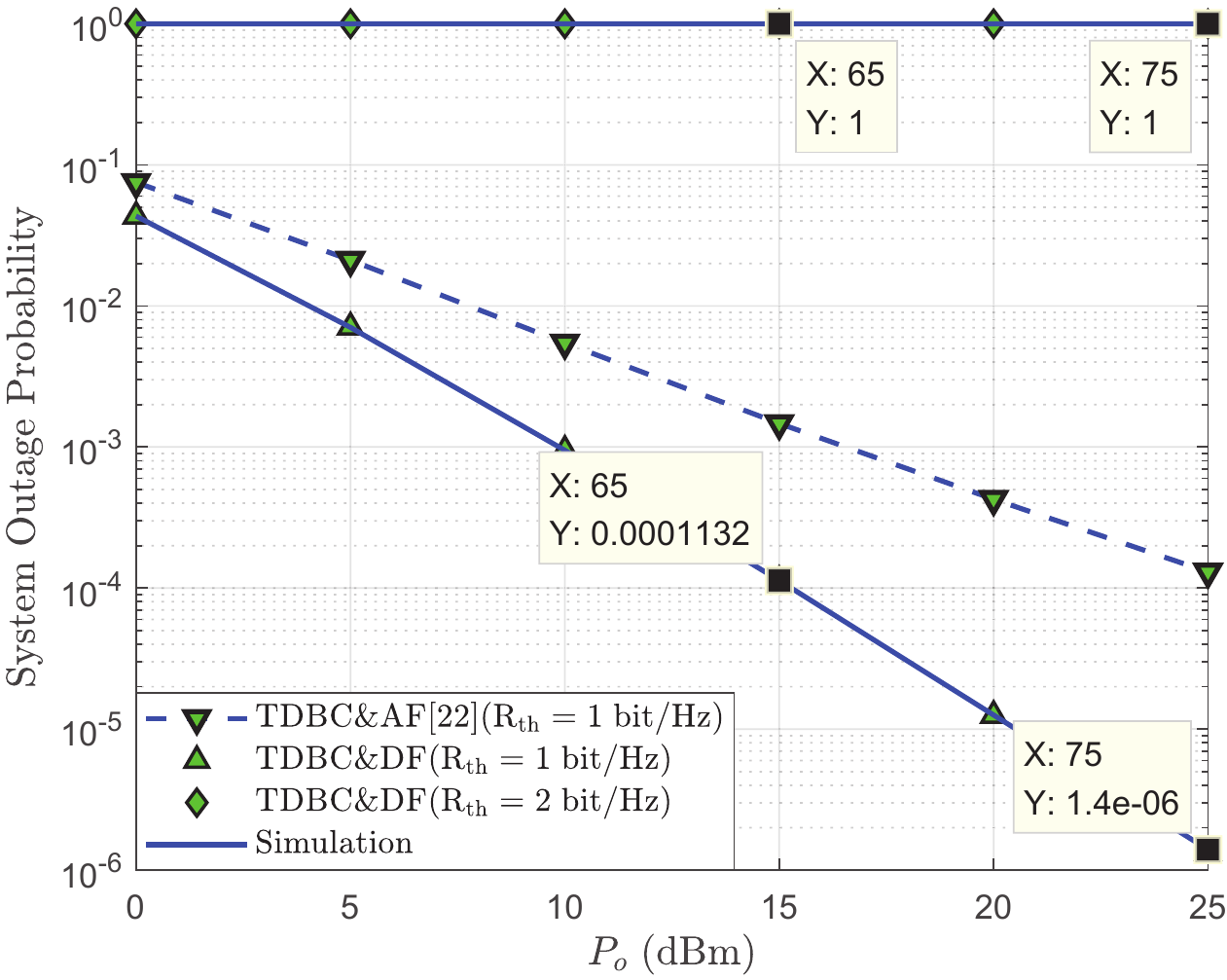} \\
  \caption{The derived diversity gain $d$ vs. the transmit power of each terminal $P_o$, in comparison with simulation results, where $k_{ave}=0.1$ and $\{m_a, m_b, m_d\}=\{2, 1, 1\}$.}
\end{figure}

Fig. 7 verifies the derived diversity gains in all two cases of the \eqref{131}. Given two points ($x_1$, $y_1$) and ($x_2$, $y_2$) in the \lq\lq$\bigtriangleup$\rq\rq or \lq\lq$\diamond$\rq\rq marked curve in Fig. 7, we can calculate the slope of the curve as -2 or 0 by $\frac{{\log \left( {{y_1}} \right) - \log \left( {{y_2}} \right)}}{{\log \left( {{{10}^{{x_1}/10}}} \right) - \log \left( {{{10}^{{x_2}/10}}} \right)}}$, which is consistent with the derived diversity gain in Section 3.2. Additionally, as presented in this figure, when the target transmission rate is below the OSC threshold, both the direct and relaying links of the considered network jointly contribute to the diversity gain. However, the achievable diversity gain of the TDBC with AF strategy is only determined by the direct link. This is due to the relay node using AF strategy amplifies the distortion noises caused HIs, which in turn results in that the relaying links have no contribution to the diversity gain.

\begin{figure}[!t]
  \centering
  \includegraphics[width=0.45\textwidth]{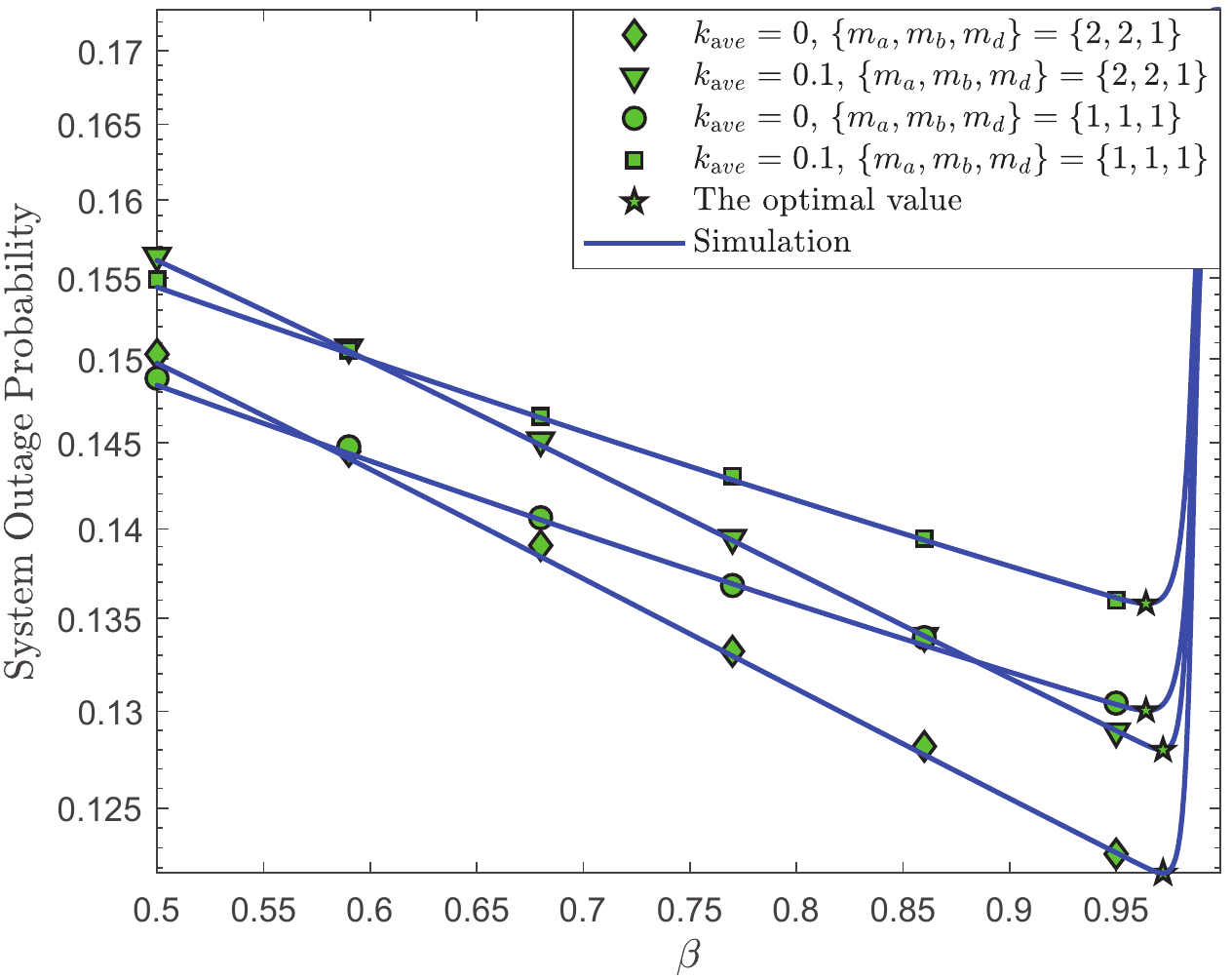} \\
  \caption{The system outage probability ${\mathbb{P}_{out}}$ vs. the PS ratio $\beta$, where $\mathbb{R}_{th}=1$ bit/Hz.}
\end{figure}

Fig. 8 depicts the system outage probability versus the PS ratio $\beta$. As presented in this figure, when $\beta$ increases, all curves decrease first and then increases.
%As presented in this figure, the system outage probability decreases first and then increases as the increase of $\beta$, viz., there exists a optimal $\beta$ to minimize the system outage probability.
Such variation trend is owing to the following fact. According to \eqref{6}, the harvested energy at $R$ gradually increases as $\beta$ increases. Meanwhile, the increase of $\beta$ also reduces the portion of the received signal used for ID at $R$. When $\beta$ increases from zero to the optimal value, the increase of the transmit power at $R$ dominantly enhance the system outage performance. However, when $\beta$ exceeds the optimal value and further increases, it is hard for $R$ to decode information from two terminals, which in turn increases the system outage probability.
%further increase in $\beta$ leads to a difficult information decoding at $R$, which in turn increases the system outage probability.
Additionally, it can be observed that, for the fixed $k_{ave}$, the optimal $\beta$ increases as shape parameters of the relaying links increase. This is because $R$ needs fewer portion of the received signal used for ID as the channel quality of the relaying links improves.

\begin{figure}[!t]
  \centering
  \includegraphics[width=0.45\textwidth]{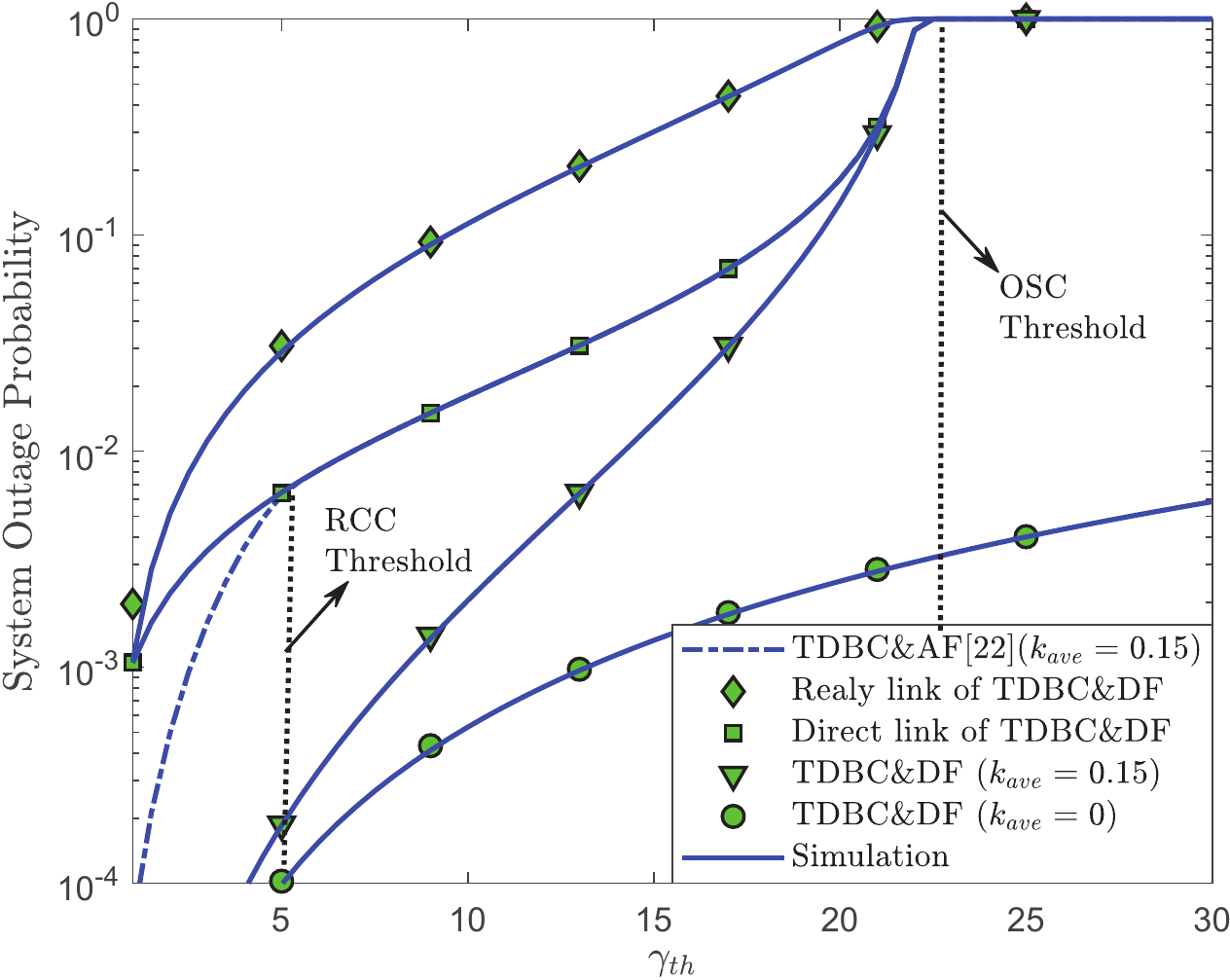} \\
  \caption{The system outage probability ${\mathbb{P}_{out}}$ vs. the SNDR threshold $\gamma_{th}$, where $P_o=10$ dBm and $\{m_a, m_b, m_d\}=\{2, 2, 1\}$.}
\end{figure}

Fig. 9 shows the system outage probability as a function of the SNDR threshold $\gamma_{th}$ using either HIs ($k_{ave}=0.15$) or ideal hardware ($k_{ave}=0$) assumption.
%Fig. 4 studies the impacts of $\gamma_{th}$ on the OSC with different HIs levels.
%%Herein, we set $\beta=0.8$ and $\rho=50$ dB.
%Let's take $k_{ave}=0.15$ as an example, for this, the OSC effects occur at the SNDRs threshold $\gamma_{th} \approx 22.2$ dB (${\gamma _{th}} \ge \frac{1}{{k_1^2 + k_2^2}}$), respectively.
From this figure, it can be observed that when $\gamma_{th}$ is below $22.2$, i.e., the OSC threshold $\frac{1}{{k_1^2 + k_2^2}}$, the system outage performance of our studied network under HIs relies on both the direct and relaying links. Nevertheless, when $\gamma_{th}$ goes beyond the OSC threshold, the overall system ceases. These results coincide with the conclusions in Remark 1.
%Referring to the analysis in section III-B and Fig. 8, we can be sure that when $\gamma_{th}$ is below the OSC threshold $\gamma_{th}\geq 22.2$ dB (${\gamma _{th}} = \frac{1}{{k_1^2 + k_2^2}}$), the system outage performance depends on both the relaying and direct links. However, when $\gamma_{th}$ exceeds the OSC threshold, the overall system is in outage.
Moreover, the impact of HIs is negligible for a low SNDR threshold, but it becomes very severe with increasing $\gamma_{th}$.
%the system outage performance for the considered network is only slightly aggravated by HIs at low SNDR threshold. However, this behavior is very different as $\gamma_{th}$  increase.
Note that, different from the studied network, the SWIPT based AF TWRN suffers from not only the OSC effect but also the RCC effect. More specifically, the RCC prevents the relaying link from carrying out cooperative communication, while the OSC puts  the overall system in outage. In addition, we also observe that the relaying links under the AF strategy are more sensitive to HIs than those under the DF strategy.
\begin{figure}[!t]
  \centering
  \includegraphics[width=0.45\textwidth]{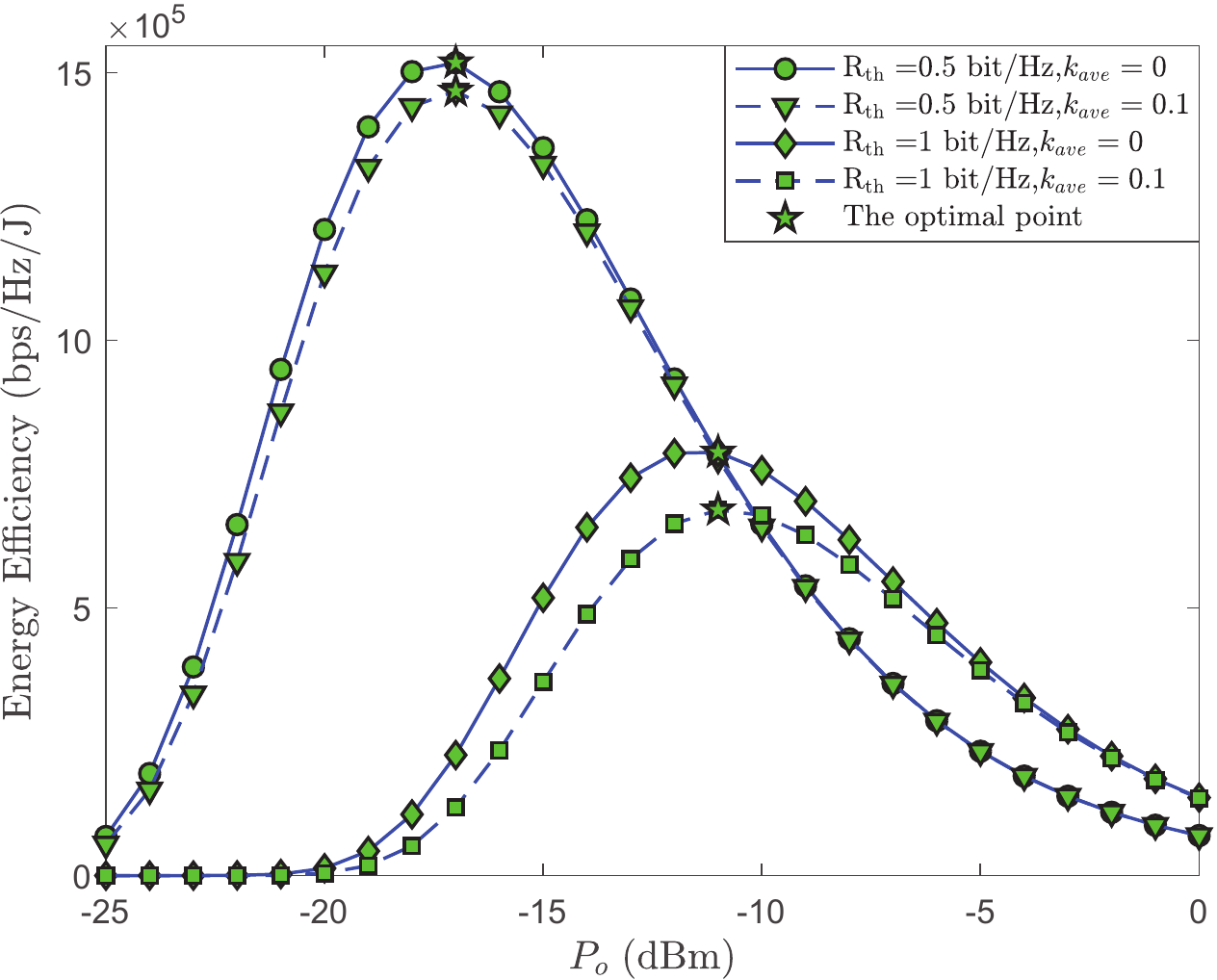} \\
  \caption{Energy efficiency vs. the transmit power of each terminal $P_o$, where $\{m_a, m_b, m_d\}=\{2, 2, 1\}$.}
\end{figure}

%Moreover, as shown in figure, the gap between the ideal hardware case ($k_{ave}=0$) and impairment cases ($k_{ave}=0.15$) is relatively small at low $\gamma_{th}$ region but other can be very large. Therefore, we can conclude that HIs are detrimental for system to keep with its outage performance standards, specially in high-data-rate applications.
Fig. 10 studies the influence of transmit power of each terminal $P_o$ on the energy efficiency
%illustrates the energy efficiency against input SNR $\rho$
to get some insights about the utilization of the available energy.
Specially, the energy efficiency is defined as $EE = \frac{{{\mathbb{R}_{th}}\left( {1 - {{\rm{\mathbb{P}}}_{out}}} \right)}}{{2T{P_o}/3}}$, where ${{\rm{\mathbb{P}}}_{out}}$ can be obtained from \eqref{16} \cite{8249542}.
From this figure, it can be observed that all curves exist the optimal $P_o$ maximizing the energy efficiency. In addition, we can also note that the achievable energy efficiency is low at high $P_o$ region. This is due to the
achieved system outage performance is much lower than the consumed energy at high transmit power region.
%consumed energy is higher than the achieved system outage performance.
Furthermore, for a fixed transmission rate, the energy efficiency under practical case ($k_{ave}=0.1$) is much lower than that under ideal case ($k_{ave}=0$). This is because the HIs impose an undesirable influence on the system outage performance of our considered network.

%increase the system outage probability of the considered network.

%Fig. 5 plots the energy efficiency against input SNR with different data rates and HIs levels. The energy efficiency is computed as $EE = \frac{{T\left( {1 - {\mathbb{P}_{\rm out}}} \right){\rm R_{th}}}}{{2T{P_o}/3}} = \frac{{3\left( {1 - {\mathbb{P}_{\rm out}}} \right){\rm R_{th}}}}{{2{P_o}}}$ \cite{8851300}. One can see that as the increase of SNR, the energy efficiency for all curves firstly increases and then decreases. This indicates that  an optimal SNR exists to maximize the energy efficiency, and   that the selection of appropriate transmit power at terminals is crucial to balance between spectral efficiency and energy efficiency. From this figure, we can also observe that, for a given data rate, the optimal energy efficiency decreases as $k_{ave}$ increases. This is duo to the detrimental impact caused by HIs on system outage performance is more serious with the increase of $k_{ave}$, which further degrades the energy efficiency. Similarly, for a fixed $k_{ave}$, the optimal energy efficiency decreases with data rate increases. %Lastly, note that the energy efficiency for all cases is lowest in high SNR region. This follows the fact that compared with the achieved system outage capacity, the consumed energy is much higher in a high SNR case.

\section{Conclusion}
In this work, we have analyzed the system outage probability and the diversity gain of a SWIPT based two-way DF relaying, in which HIs at all transceivers are taken into account. In particular, under i.n.i.d. Nakagami-$m$ fading channels, the closed-form expression for the system outage probability has been obtained. Based the derived expression, we have identified the OSC effect and obtained the achievable diversity gain of our studied network. Our analytical results have revealed the following fact, i.e.,  when transmission rate goes beyond the OSC threshold, the overall system is in outage and the diversity gain is zero; otherwise, both the direct and relaying links contribute to enhancing the system outage performance and the resulting diversity gain is the sum of the shape parameter of the direct link and the smaller shape parameter of the terminal-to-relay links, viz., $d=m_d+\min \left( {{m_a},{m_b}} \right)$. In addition, numerical results have provided some insights about the effect of multifarious parameters on system outage performance as well as the performance difference between DF and AF strategies. Based on the results, we have provided guideline on how terminals use energy to balance the energy utilization and spectral utilization.

\appendices
\begin{appendices}
\section{}
In order to obtain the closed-form expression for $\mathbb{P}_1$, we divide into the following two cases in terms of the range of ${\gamma _{th}}$.
%Based on the expression $\mathbb{P}_1$ of \eqref{11}, we divide into the following two cases to obtain the corresponding closed-form expression.
 %According to the expression for $\mathbb{P}_1$, we derive its closed-form expression in the following two cases.

\emph{Case 1:} When ${\gamma _{th}} \ge \frac{1}{{k_1^2 + k_2^2}}$ holds, the inequality of $\mathbb{P}_1$, ${\left( {1 - \left( {k_1^2 + k_2^2} \right){\gamma _{th}}} \right)\rho Z < {\gamma _{th}}}$, is always true no matter what value random variable $Z$ $(Z>0)$ takes. Therefore, the corresponding probability equals one, viz., $\mathbb{P}_1=1$.

\emph{Case 2:} When ${\gamma _{th}} < \frac{1}{{k_1^2 + k_2^2}}$ is satisfies, $\mathbb{P}_1$ can be re-expressed as $\mathbb{P}_1^1$, given by
\setcounter{equation}{29}
\begin{align}\label{34}
  {\mathbb{P}_1^1} &= \Pr \left( {Z < \frac{{{\gamma _{th}}}}{\rho{\left( {1 - {\gamma _{th}}\left( {k_1^2 + k_2^2} \right)} \right)}}} \right)\nonumber\\
  &= \int_0^{\frac{{{\gamma _{th}}}}{\rho{\left( {1 - {\gamma _{th}}\left( {k_1^2 + k_2^2} \right)} \right)}}} {\frac{1}{{\Gamma \left( {{m_d}} \right)\theta _d^{{m_d}}}}} {z^{{m_d} - 1}}{e^{ - \frac{z}{{{\theta _d}}}}}dz \tag{A.1},
\end{align}
where $m_d$ and ${\theta _d} = \frac{{{\Omega _d}}}{{{m_d}}}$ are the shape and scale parameters of the gamma random variable $Z$ respectively. According to [29, eq.(3.381.1)] and [29, eq.(8.352.6)], we can solve the integration of \eqref{34}, and then obtain the closed-form expression for ${\mathbb{P}_{1}^1}$ as given in \eqref{13}.

\section{}
Based on the above analysis of $\mathbb{P}_1$, we can also derive the closed-form expression for $\mathbb{P}_2$ in terms of the following two cases.
%Similar to the derivation of $\mathbb{P}_1$, the derivation of $\mathbb{P}_2$ can also be split into the following two cases.
%Referring to the derivation of $\mathbb{P}_1$, we can also derive the closed-form expression for $\mathbb{P}_2$ in following two cases.

\emph{Case 1:} When ${\gamma _{th}} \ge \frac{1}{{k_1^2 + k_2^2}}$ is satisfied,  all the inequalities of $\mathbb{P}_2$ are never true. Therefore, the corresponding probability is equal to zero, i.e., $\mathbb{P}_2=1$.

\emph{Case 2:} When ${\gamma _{th}} < \frac{1}{{k_1^2 + k_2^2}}$ holds, $\mathbb{P}_2$ can be rewritten as $\mathbb{P}_2 = \Pr \{ X > {\Delta _1},Y > {\Delta _1},X >  - Y + {\Delta _2}/Y,Y >  - X + {\Delta _2}/X\} $, where ${\Delta_1} = \frac{{{\gamma _{th}}}}{{\left( {1 - \beta } \right)\left( {1 - \left( {k_1^2 + k_2^2} \right){\gamma _{th}}} \right)\rho }}$ and ${\Delta_2} = \frac{{{\gamma _{th}}}}{{\eta \beta \left( {1 - \left( {k_1^2 + k_2^2} \right){\gamma _{th}}} \right)\rho }}$. Obviously, there is the high correlation among the inequalities in $\mathbb{P}_2$, which is the main obstacle in deriving $\mathbb{P}_2$.
%it is difficult to directly obtain the corresponding closed-form expression.
%Due to the high correlation of inequalities in $\mathbb{P}_2$, it is difficult to obtain the closed-form expression directly.
%In order to obtain its solution, we first analyze and determine the integral region. According to the integral region, we further obtain the closed-form expression for $\mathbb{P}_2$ by calculating the integral value.
Assume that $X=x$ and $Y=y$. From the expression of $\mathbb{P}_2$ in \eqref{11}, the integral region of $\mathbb{P}_2$ is bounded by four curves, which are ${l_1}:x = {\Delta _1}$, ${l_2}:y = {\Delta _1}$, ${l_3}:x =  - y + {\Delta _2}/y$ and ${l_4}:y =  - x + {\Delta _2}/x$. After straightforward  mathematical manipulations for the above four curves, we can obtain several conclusions as follows:

1) curves $l_3(y)$ and $l_4(x)$ are the monotonically decreasing functions with respect of $y$ and $x$ respectively, where $x,y \in \left( {0, + \infty } \right)$.

%for curve $l_3$ ($l_4$), $x$ (or $y$)  decreases monotonically when $y$ (or $x$) increases from 0 to $+\infty$;

2) $\mathop {\lim }\limits_{x \to 0} \left( { - x + {\Delta _2}/x} \right) =  + \infty $ and $\mathop {\lim }\limits_{y \to 0} \left( { - y + {\Delta _2}/y} \right) =  + \infty$;

3) the zero point of curve $l_3$ ($l_4$) at y(x)-axis is $y = \sqrt {{\Delta _2}}$ $\left(x = \sqrt {{\Delta _2}} \right)$;

4) curves $l_3$ and $l_4$ are symmetric about $l_5: y=x$;

5) the intersection between curves $l_3$ and $l_4$ is $\left( {\sqrt {{\Delta _2}/2} ,\sqrt {{\Delta _2}/2} } \right)$ in the first quadrant;

6) the intersection between curves $l_3$ ($l_4$) and $l_5$ is $\left( {\sqrt {{\Delta _2}/2} ,\sqrt {{\Delta _2}/2} } \right)$ in the first quadrant;

Based on the above conclusions, there are two cases for the integral region of $\mathbb{P}_2$, discussed as follows.

\begin{figure}[!t]
  \centering
  \includegraphics[width=0.34\textwidth]{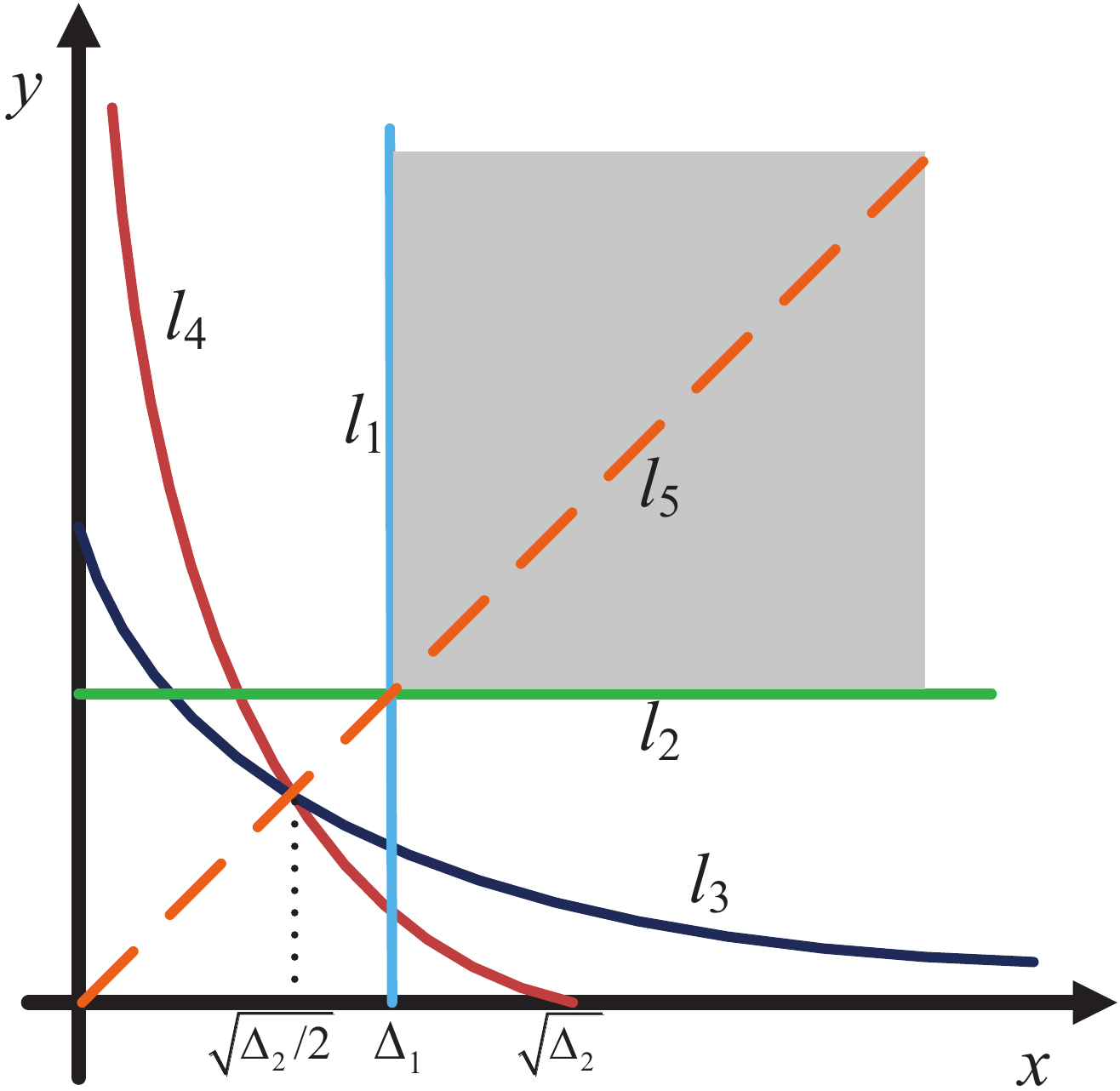} \\
  \caption{The integral region for $\mathbb{P}_2$, where ${\Delta _{\rm{1}}} \ge \sqrt {{\Delta _{\rm{2}}}{\rm{/2}}}$.}
\end{figure}

\emph{i):} When ${\Delta _{\rm{1}}} \ge \sqrt {{\Delta _{\rm{2}}}{\rm{/2}}}$ holds, the integral region for $\mathbb{P}_2$ can be shown as the shadow area in Fig. 11. Thus, $\mathbb{P}_2$ can be expressed as ${\mathbb{P}_{2}^1}$, given by
\begin{align}
{\mathbb{P}_2^1} =& \underbrace {\int_{{\Delta _1}}^\infty  {\int_{{\Delta _1}}^x {{f_Y}\left( y \right){f_X}\left( x \right)dydx} } }_{{\mathbb{P}_{2,1}^1} }\nonumber\\
&+ \underbrace {\int_{{\Delta _1}}^\infty  {\int_{{\Delta _1}}^y {{f_X}\left( x \right){f_Y}\left( y \right)dxdy} } }_{{\mathbb{P}_{2,2}^1}}.\tag{B.1}
\end{align}

The first term of (B.1), ${\mathbb{P}_{2,1}^1}$, can be calculated as
\begin{align}
&\mathbb{P}_{2,1}^1 =\int_{{\Delta _1}}^\infty  {\int_{{\Delta _1}}^x {\frac{1}{{\Gamma \left( {{m_a}} \right)\Gamma \left( {{m_b}} \right)\theta _a^{{m_a}}\theta _b^{{m_b}}}}} }\nonumber\\
&\;\;\;\;\;\;\;\;\;\;\times {x^{{m_a} - 1}}{y^{{m_b} - 1}}{e^{ - \frac{x}{{{\theta _a}}} - \frac{y}{{{\theta _b}}}}}dydx\nonumber\\
&=\! \underbrace {\int_{{\Delta _1}}^\infty  {\frac{1}{{\Gamma \left(\! {{m_a}}\! \right)\Gamma \left( \!{{m_b}}\! \right)\theta _a^{{m_a}}}}{x^{{m_a}\! - \!1}}{e^{\! -\! \frac{x}{{{\theta _a}}}}}} \gamma \left(\! {{m_b},x/{\theta _b}} \!\right)dx}_{{\Xi _1}}\nonumber\\
&-\!\underbrace {\int_{{\Delta _1}}^\infty  {\frac{1}{{\Gamma \left(\! {{m_a}}\! \right)\Gamma \left(\! {{m_b}}\! \right)\theta _a^{{m_a}}}}{x^{{m_a} \!- \!1}}{e^{\! -\! \frac{x}{{{\theta _a}}}}}} \gamma \left(\! {{m_b},{\Delta _1}\!/\!{\theta _b}}\! \right)dx}_{{\Xi _2}}.\tag{B.2}
\end{align}

Using [29, 8.352.6], the first term of $\mathbb{P}_{2,1}^1$, ${\Xi _1}$, can be rewritten as
\begin{align}
&{\Xi _1} = \int_{{\Delta _1}}^\infty  {\frac{1}{{\Gamma \left( {{m_a}} \right)\theta _a^{{m_a}}}}{x^{{m_a} - 1}}{e^{ - \frac{x}{{{\theta _a}}}}}} dx - \frac{1}{{\Gamma \left( {{m_a}} \right)\theta _a^{{m_a}}}}\nonumber\\
&\!-\! \frac{1}{{\Gamma \left(\! {{m_a}} \!\right)\theta _a^{{m_a}}}}\sum\limits_{l\! = \!0}^{{m_b}\! -\! 1} {\frac{1}{{l!}}{{\left(\! {\frac{1}{{{\theta _b}}}}\! \right)}^l}} \int_{{\Delta _1}}^\infty  {{x^{l \!+ \!{m_a}\! -\! 1}}{e^{ \!- \!\left( \!{\frac{1}{{{\theta _a}}}\! + \!\frac{1}{{{\theta _b}}}}\! \right)x}}}\!dx.\tag{B.3}
\end{align}

According to  [29, 3.381.3], ${\Xi _1}$ can be calculated as
\begin{align}
&{\Xi _1} = \frac{1}{{\Gamma \left( {{m_a}} \right)}}\Gamma \left( {{m_a},{\Delta _1}/{\theta _a}} \right) - \frac{1}{{\Gamma \left( {{m_a}} \right)\theta _a^{{m_a}}}}\sum\limits_{l = 0}^{{m_b} - 1} {\frac{1}{{l!}}}\nonumber\\
&\!\times\! {\left( \!{\frac{1}{{{\theta _b}}}} \!\right)^l}{\left(\! {\frac{1}{{{\theta _a}}}\! +\! \frac{1}{{{\theta _b}}}} \!\right)^{ \!- \!\left( {l \!+\! {m_a}} \right)}}\Gamma \left(\! {l\! + \!{m_a},{\Delta _1}\left( {\frac{1}{{{\theta _a}}}\! +\! \frac{1}{{{\theta _b}}}}\! \right)}\! \right).\tag{B.4}
\end{align}

The second term of $\mathbb{P}_{2,1}^1$, ${\Xi _2}$, can be calculated as
\begin{align}
{\Xi _2} \!= \!\frac{1}{{\Gamma \left(\! {{m_b}}\! \right)}}\gamma \left(\! {{m_b},\!{\Delta _1}/{\theta _b}}\! \right)\left(\! {1\! -\! \frac{1}{{\Gamma \left(\! {{m_a}} \!\right)}}\gamma \left(\! {{m_a},\!{\Delta _1}/{\theta _a}} \!\right)}\! \right).\tag{B.5}
\end{align}

Referring to the analysis of ${\mathbb{P}_{2,1}^1}$, we can also derive ${\mathbb{P}_{2,2}^1}$ in a closed-form.
%obtain the closed-form expression for ${\mathbb{P}_{2,2}^1}$.
Based on the above derivations, the closed-form expression for ${\mathbb{P}_{2}^1}$ can be obtained as \eqref{19}.

\begin{figure*}[!t]
\normalsize
\begin{align}\label{20}
&\mathbb{P}_{2,1}^2 =\nonumber\\
& \underbrace {\frac{1}{{\Gamma \left( {{m_a}} \right)\theta _a^{{m_a}}}}\sum\limits_{l = 0}^{{m_b} - 1} {\frac{1}{{l!}}} {{\left( {\frac{1}{{{\theta _b}}}} \right)}^l}\int_0^{\Phi  - \sqrt {{\Delta _2}/2} } {{{\left( {x + \sqrt {{\Delta _2}/2} } \right)}^{{m_a} - 1}}{Q^l}\left( {x + \sqrt {{\Delta _2}/2} } \right){e^{ - \frac{{x + \sqrt {{\Delta _2}/2} }}{{{\theta _a}}} - \frac{{Q\left( {x + \sqrt {{\Delta _2}/2} } \right)}}{{{\theta _b}}}}}} dx}_{{\Xi _3}}\nonumber\\
&- \underbrace {\frac{1}{{\Gamma \left( {{m_a}} \right)\theta _a^{{m_a}}}}\sum\limits_{l = 0}^{{m_b} - 1} {\frac{1}{{l!}}} {{\left( {\frac{1}{{{\theta _b}}}} \right)}^l}\int_0^{\Phi  - \sqrt {{\Delta _2}/2} } {{{\left( {x + \sqrt {{\Delta _2}/2} } \right)}^{{m_a} + l - 1}}{e^{ - \frac{{x + \sqrt {{\Delta _2}/2} }}{{{\theta _a}}} - \frac{{x + \sqrt {{\Delta _2}/2} }}{{{\theta _b}}}}}} dx}_{{\Xi _4}}.\tag{B.8}
\end{align}
\setcounter{equation}{22}
\hrulefill
%\vspace*{-5pt}
\end{figure*}

\begin{figure}[!t]
  \centering
  \includegraphics[width=0.34\textwidth]{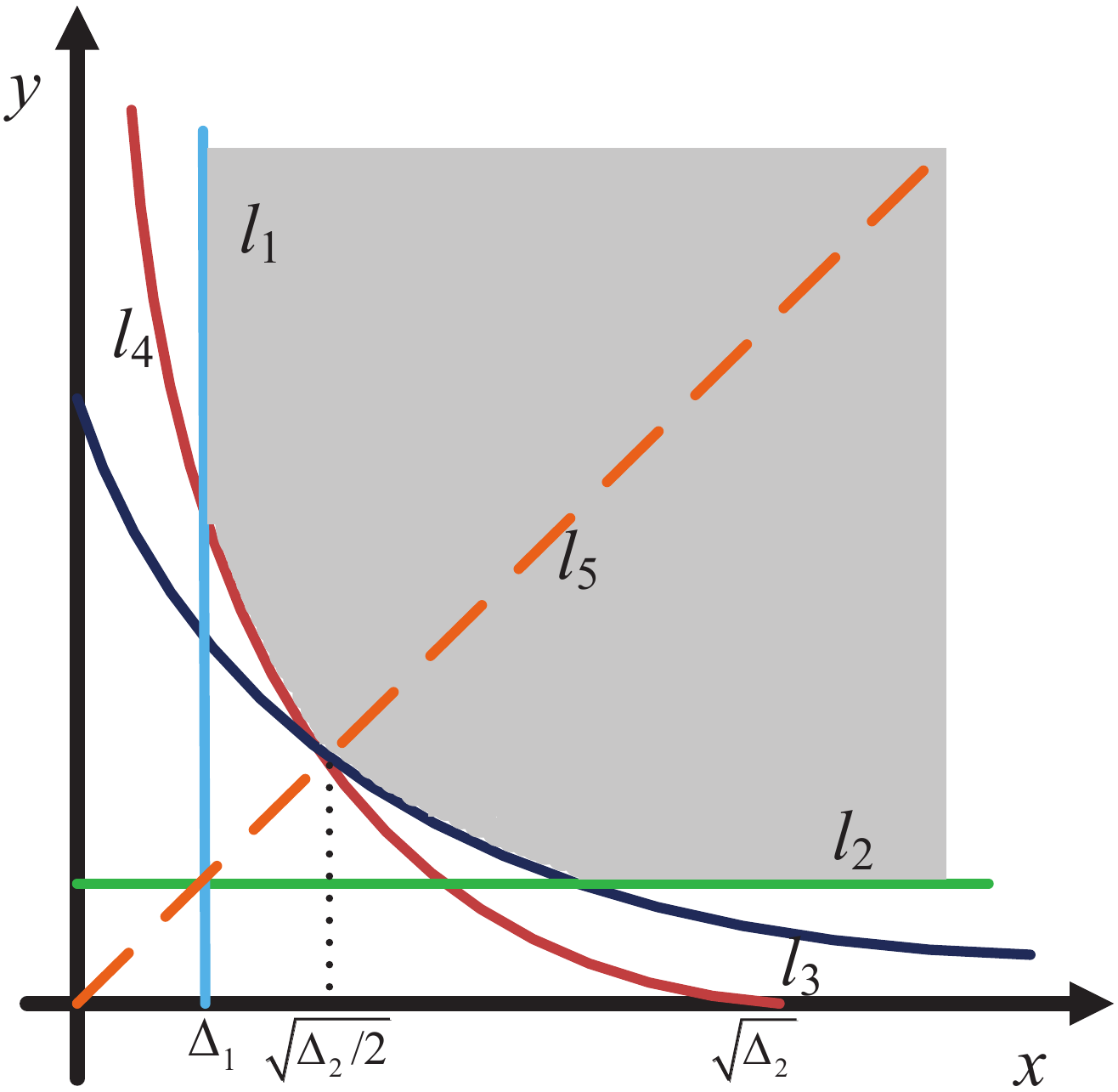} \\
  \caption{The integral region for $\mathbb{P}_2$, where ${\Delta _{\rm{1}}} < \sqrt {{\Delta _{\rm{2}}}{\rm{/2}}}$.}
\end{figure}

\emph{ii):} When ${\Delta _{\rm{1}}} < \sqrt {{\Delta _{\rm{2}}}{\rm{/2}}}$ holds, the integral region for $\mathbb{P}_2$ can be shown as the shadow area in Fig. 12 at the top of the next page. Hence, $\mathbb{P}_2$ can be expressed as ${\mathbb{P}_{2}^2}$, given by
\begin{align}
%\mathbb{P}_2^2 =& \underbrace {\int_{\sqrt {{\Delta _2}/2} }^\Phi  {\int_{Q\left( x \right)}^x {{f_Y}\left( y \right){f_X}\left( x \right)dydx} } }_{\mathbb{P}_{2,1}^2}\nonumber\\
%&+ \underbrace {\int_\Phi ^\infty  {\int_{{\Delta _1}}^x {{f_Y}\left( y \right){f_X}\left( x \right)dydx} } }_{\mathbb{P}_{2,2}^2}\nonumber\\
\mathbb{P}_2^2 =& \underbrace {\int_{\sqrt {{\Delta _2}/2} }^\Phi  {\int_{Q\left( x \right)}^x {{f_Y}\left( y \right){f_X}\left( x \right)dydx} } }_{\mathbb{P}_{2,1}^2}\nonumber\\
&+ \underbrace {\int_\Phi ^\infty  {\int_{{\Delta _1}}^x {{f_Y}\left( y \right){f_X}\left( x \right)dydx} } }_{\mathbb{P}_{2,2}^2}\nonumber\\
&+ \underbrace {\int_{\sqrt {{\Delta _2}/2} }^\Phi  {\int_{Q\left( y \right)}^y {{f_X}\left( x \right){f_Y}\left( y \right)dxdy} } }_{\mathbb{P}_{2,3}^2}\nonumber\\
&+ \underbrace {\int_\Phi ^\infty  {\int_{{\Delta _1}}^y {{f_X}\left( x \right){f_Y}\left( y \right)dx} } dy}_{\mathbb{P}_{2,4}^2},\tag{B.6}
\end{align}
where $\Phi  =  - {\Delta _1} + {\Delta _2}/{\Delta _1}$, $Q\left( x \right) = \left( { - x + \sqrt {{x^2} + 4{\Delta _2}} } \right)/2$ and $Q\left( y \right) = \left( { - y + \sqrt {{y^2} + 4{\Delta _2}} } \right)/2$.

By means of variable substitution, $\mathbb{P}_{2,1}^2$ can be expressed as
\begin{align}
&\mathbb{P}_{2,1}^2 =\int_0^{\Phi  - \sqrt {{\Delta _2}/2} } {\frac{1}{{\Gamma \left( {{m_a}} \right)\Gamma \left( {{m_b}} \right)\theta _a^{{m_a}}}}}\nonumber\\
&\! \times\!{{\left(\! {x\! + \!\sqrt {{\Delta _2}/2} }\! \right)}^{{m_a}\! - \!1}}{e^{ \!-\! \frac{{x \!+\! \sqrt {{\Delta _2}/2} }}{{{\theta _a}}}}}\Bigg(\gamma \left(\! {{m_b},\frac{{x\! + \!\sqrt {{\Delta _2}/2} }}{{{\theta _b}}}} \!\right)\nonumber\\
&- \gamma \left( {{m_b},\frac{{Q\left( {x + \sqrt {{\Delta _2}/2} } \right)}}{{{\theta _b}}}} \right)\Bigg)dx.\tag{B.7}
\end{align}
According to [29, 8.352.6], $\mathbb{P}_{2,1}^2$ can be rewritten as \eqref{20}, as presented at the top of the next page.
Due to the high complexity for the integral in ${\Xi _3}$, we adopt the Guassian-Chebyshev quadrature$^7$\footnotetext[7]{Owing to the sufficient accuracy even with very few terms, we adopt the Gaussian-Chebyshev quadrature instead of other approximation approaches.
Note that such a method has been widely adopted in existing works \cite{8633928,8613860,8576644,8026173,9055221}.} to approximate it.
%Since the closed-form expression for the required integral in ${\Xi _3}$ can not be obtained directly, we adopt the Guassian-Chebyshev quadrature\footnote{Owing to the sufficient accuracy with very few terms, we adopt the Gaussian-Chebyshev quadrature instead of other approximation approaches. Note that such method has been widely adopted in existing works \cite{8633928,8613860,8576644,8026173,9055221}.} to approximate it.
Thus, the first term of $\mathbb{P}_{2,1}^2$, ${\Xi _3}$, can be calculated as
\begin{align}
{\Xi _3} = &\frac{{\pi \left( {\Phi  - \sqrt {{\Delta _2}/2} } \right)}}{{2N\Gamma \left( {{m_a}} \right)\theta _a^{{m_a}}}}\sum\limits_{l = 0}^{{m_b} - 1} {\sum\limits_{n = 1}^N {\sqrt {1 - v_n^2} \frac{1}{{l!}}{{\left( {\frac{1}{{{\theta _b}}}} \right)}^l}} }\nonumber\\
&\times {\left( {k_n^{\rq} + \sqrt {{\Delta _2}/2} } \right)^{{m_a} - 1}}{Q^l}\left( {k_n^{\rq} + \sqrt {{\Delta _2}/2} } \right)\nonumber\\
&\times {e^{ - \frac{{k_n^{\rq} + \sqrt {{\Delta _2}/2} }}{{{\theta _a}}} - \frac{{Q\left( {k_n^{\rq} + \sqrt {{\Delta _2}/2} } \right)}}{{{\theta _b}}}}},\tag{B.9}
\end{align}
where ${v_n} = \cos \left( {\frac{{\left( {2n - 1} \right)\pi }}{{2N}}} \right)$, $k_n^{\rq} = \frac{{\Phi  - \sqrt {{\Delta _2}/2} }}{2}\left( {{v_n} + 1} \right)$ and $N$ is the complexity and accuracy tradeoff parameter.

Adopting variable substitution, the second term of $\mathbb{P}_{2,1}^2$, ${\Xi _4}$, can be written as
\begin{align}
&{\Xi _4}\! =\! \frac{1}{{\Gamma \left( \!{{m_a}} \!\right)\theta _a^{{m_a}}}}\sum\limits_{l\! =\! 0}^{{m_b}\! - \!1} {\frac{1}{{l!}}} {\left(\! {\frac{1}{{{\theta _b}}}} \!\right)^l}{\left(\! {\frac{{{\theta _a}{\theta _b}}}{{{\theta _a}\! +\! {\theta _b}}}}\! \right)^{{m_a}\! +\! l}}\;\;\;\;\;\;\;\;\;\;\;\;\nonumber\\
&\times {e^{ - \sqrt {{\Delta _2}/2} \left( {\frac{1}{{{\theta _a}}} + \frac{1}{{{\theta _b}}}} \right)}}\int_0^{\left( {\frac{1}{{{\theta _a}}} + \frac{1}{{{\theta _b}}}} \right)\left( {\Phi  - \sqrt {{\Delta _2}/2} } \right)}\nonumber\\
&\!\times\!{\left(\! {x\! + \!\sqrt {{\Delta _2}/2} \left(\! {\frac{1}{{{\theta _a}}} + \frac{1}{{{\theta _b}}}} \right)} \right)^{{m_a} + l - 1}}{e^{ - x}}dx.\tag{B.10}
\end{align}

Using [29, 3.382.5], ${\Xi _4}$ can be calculated as
\begin{align}
{\Xi _4} = &\frac{1}{{\Gamma \left( {{m_a}} \right)\theta _a^{{m_a}}}}\sum\limits_{l = 0}^{{m_b} - 1} {\frac{1}{{l!}}} {\left( {\frac{1}{{{\theta _b}}}} \right)^l}{\left( {\frac{{{\theta _a}{\theta _b}}}{{{\theta _a} + {\theta _b}}}} \right)^{{m_a} + l}}\nonumber\\
&\times \bigg(\gamma \left( {{m_a} + l,\Phi \left( {\frac{1}{{{\theta _a}}} + \frac{1}{{{\theta _b}}}} \right)} \right)\nonumber\\
&- \gamma \left( {{m_a} + l,\sqrt {{\Delta _2}/2} \left( {\frac{1}{{{\theta _a}}} + \frac{1}{{{\theta _b}}}} \right)} \right)\bigg).\tag{B.11}
\end{align}

On the basis of the derivation of ${\mathbb{P}_{2}^1}$, $\mathbb{P}_{2,2}^2$ can be calculated as
\begin{align}
&\mathbb{P}_{2,2}^2 = \frac{1}{{\Gamma \left( {{m_a}} \right)}}\Gamma \left( {{m_a},\Phi /{\theta _a}} \right) - \frac{1}{{\Gamma \left( {{m_a}} \right)\theta _a^{{m_a}}}}\sum\limits_{l = 0}^{{m_b} - 1} {\frac{1}{{l!}}}\notag
\end{align}
\begin{align}
%&\mathbb{P}_{2,2}^2 = \frac{1}{{\Gamma \left( {{m_a}} \right)}}\Gamma \left( {{m_a},\Phi /{\theta _a}} \right) - \frac{1}{{\Gamma \left( {{m_a}} \right)\theta _a^{{m_a}}}}\sum\limits_{l = 0}^{{m_b} - 1} {\frac{1}{{l!}}}\nonumber\\
&\!\times\! {\left(\! {\frac{1}{{{\theta _b}}}}\! \right)^l}{\left(\! {\frac{1}{{{\theta _a}}}\! +\! \frac{1}{{{\theta _b}}}} \!\right)^{ \!-\! \left( {l \!+\! {m_a}} \right)}}\Gamma \left( {l\! +\!{m_a},\Phi \left( {\frac{1}{{{\theta _a}}}\! + \!\frac{1}{{{\theta _b}}}} \right)} \right)\nonumber\\
&\!-\! \frac{1}{{\Gamma \left( {{m_b}} \right)}}\gamma \left( {{m_b},{\Delta _1}/{\theta _b}} \right)\left( {1\! -\! \frac{1}{{\Gamma \left(\! {{m_a}} \right)}}\gamma \left(\! {{m_a},\Phi /{\theta _a}}\! \right)}\! \right).\tag{B.12}
\end{align}

In addition, referring to the analyses of ${\mathbb{P}_{2,1}^2}$ and ${\mathbb{P}_{2,2}^2}$, we can also derive ${\mathbb{P}_{2,3}^2}$ and ${\mathbb{P}_{2,4}^2}$ in closed-forms respectively. Based on the above derivation, the closed-form expression for ${\mathbb{P}_{2}^2}$ can be obtained as \eqref{21}.

\section{}
According to \eqref{16}, the diversity gain can be calculated respectively in the two cases as follows.

\emph{i)}: When ${\gamma _{th}} \ge \frac{1}{{k_1^2 + k_2^2}}$ is satisfied, $d =  - \mathop {\lim }\limits_{\rho  \to \infty } \frac{{\log \left( {\rm{1}} \right)}}{{\log \left( \rho  \right)}}{\rm{ = 0}}$;

\emph{ii)}: When ${\gamma _{th}} < \frac{1}{{k_1^2 + k_2^2}}$ holds, the diversity gain can be expressed as
\begin{align}\label{36}
d =  - \underbrace {\mathop {\lim }\limits_{\rho  \to \infty } \frac{{\log \left( {\mathbb{P}_{1}^1} \right)}}{{\log \left( \rho  \right)}}}_{{\Xi _5}} - \underbrace {\mathop {\lim }\limits_{\rho  \to \infty } \frac{{\log \left(1- {\mathbb{P}_{2}^*} \right)}}{{\log \left( \rho  \right)}}}_{{\Xi _6}}.\tag{C.1}
\end{align}

Using [29, 8.352.6] and $\gamma \left( {n,z} \right)\mathop  \simeq \limits^{z \to 0} {z^n}/n$ \cite{8405612}, the first term of \eqref{36}, ${\Xi _5}$, can be calculated as
\begin{align}
{\Xi _5} &= \mathop {\lim }\limits_{\rho  \to \infty } \frac{{\log \left( {{{\left( {\frac{{{\gamma _{th}}}}{{{\theta _d}\rho \left( {1 - {\gamma _{th}}\left( {k_1^2 + k_2^2} \right)} \right)}}} \right)}^{{m_d}}}/{m_d}!} \right)}}{{\log \left( \rho  \right)}}\nonumber\\
&=  - {m_d}.\tag{C.2}
\end{align}

According to the discussion in Appendix B, as the input SNR $\rho$ approaches infinity, $\mathbb{P}_{2}^* \approx \mathbb{P}_{2}^1$ holds. Moreover, a careful observation of \eqref{19} reveals that the probability $\mathbb{P}_{2}^1(\rho)$ increases with $\mathbb{P}_2^1(\rho ) \propto \frac{1}{{\Gamma \left( \Lambda  \right)}}\Gamma \left( {\Lambda ,{\Delta _1}/{\theta _i}} \right)$, where $\Lambda  = \min \left( {{m_a},{m_b}} \right)$, ${\Delta_1} = \frac{{{\gamma _{th}}}}{{\left( {1 - \beta } \right)\left( {1 - \left( {k_1^2 + k_2^2} \right){\gamma _{th}}} \right)\rho }}$ and when $\Lambda  = {m_a}\left( {{m_b}} \right)$, $i=a (b)$. Hence, the second term of \eqref{36}, ${\Xi _6}$, can be approximated as
\begin{align}
{\Xi _6} \approx \mathop {\lim }\limits_{\rho  \to \infty } \frac{{\log \left( {1 - \frac{1}{{\Gamma \left( \Lambda  \right)}}\Gamma \left( {\Lambda ,{\Delta _1}/{\theta _i}} \right)} \right)}}{{\log \left( \rho  \right)}}.\tag{C.3}
\end{align}
Similar to the derivation of ${\Xi _5}$, we can obtain ${\Xi _6}=-\min \left( {{m_a},{m_b}} \right)$.

Based on the above discussions, the diversity gain can be obtained as \eqref{131}.
\end{appendices}
\nocite{Gradshteyn2007Table}

\ifCLASSOPTIONcaptionsoff
  \newpage
\fi
\bibliographystyle{IEEEtran}
\bibliography{refa}
\end{document}